\documentclass[twocolumn,english,aps, pra, eqsecnum, longbibliography]{revtex4-2}
\usepackage[T1]{fontenc}
\usepackage[latin9]{inputenc}
\usepackage{array}
\usepackage{booktabs}
\usepackage{bm}
\usepackage{multirow}
\usepackage{amsmath}
\usepackage{amssymb}
\usepackage{graphicx}

\makeatletter

\providecommand{\tabularnewline}{\\}

\makeatother

\usepackage{babel}
\begin{document}
\title{Validation tests of GBS quantum computers give evidence for quantum
advantage with a decoherent target }
\author{Alexander S. Dellios, Bogdan Opanchuk, Margaret D. Reid and Peter
D. Drummond}
\affiliation{Centre for Quantum Science and Technology Theory, Swinburne University
of Technology, Melbourne 3122, Australia}
\begin{abstract}
Computational validation is vital for all large-scale quantum computers.
One needs computers that are both fast and accurate. Here we apply
precise, scalable, high order statistical tests to data from large
Gaussian boson sampling (GBS) quantum computers that claim quantum
computational advantage. These tests can be used to validate the output
results for such technologies. Our method allows investigation of
accuracy as well as quantum advantage. Such issues have not been investigated
in detail before. Our highly scalable technique is also applicable
to other applications of linear bosonic networks. We utilize positive-P
phase-space simulations of grouped count probabilities (GCP) as a
fingerprint for verifying multi-mode data. This is exponentially more
efficient than other phase-space methods, due to much lower sampling
errors. We randomly generate tests from exponentially many high-order,
grouped count tests. Each of these can be efficiently measured and
simulated, providing a quantum verification method that is non-trivial
to replicate classically. We give a detailed comparison of theory
with a 144-channel GBS experiment, including grouped correlations
up to the largest order measured. We show how one can disprove faked
data, and apply this to a classical count algorithm. There are multiple
distance measures for evaluating the fidelity and computational complexity
of a distribution. We compute these and explain them. The best fit
to the data is a partly thermalized Gaussian model, which is neither
the ideal case, nor the model that gives classically computable counts.
Even with a thermalized model, discrepancies of $Z>100$ were observed
from some $\chi^{2}$ tests, indicating likely parameter estimation
errors. Total count distributions were much closer to a thermalized
quantum model than the classical model, giving evidence consistent
with quantum computational advantage for a modified target problem. 
\end{abstract}
\maketitle

\section{Introduction}

Computers of all types require validation. Yet this is nontrivial
with quantum computers which claim quantum advantage, since the outputs
are not classically computable \citep{AaronsonArkhipov2013LV,Hamilton2017PhysRevLett.119.170501,quesada2018gaussian,boulandComplexityVerificationQuantum2019,deshpandeQuantumComputationalAdvantage2022a}.
There is much recent research on this topic due to decoherence and
noise which both limits the quantum advantage obtainable, and can
lead to erroneous results \citep{zlokapa2023boundaries}. The challenge
with any classical or quantum computer is that there are exponentially
many possible logical programs to test. Verification of large gate-based
quantum computers is currently restricted to sub-classes of programs
that allow benchmarking, or else uses extrapolation to larger sizes
than can be verified directly \citep{kim2023evidence}. When problems
have random number outputs, the distributions are usually exponentially
sparse, which in standard practice requires binning the data to measure
probabilities in order to do statistical testing \citep{knuth2014art}.

Bosonic networks employed as quantum computers combine all three of
these validation challenges of hardness, exponentially many combinations,
and output sparseness. They are developing at an increasingly rapid
pace due to their high degree of scalability. These networks include
boson samplers, which utilize either nonclassical input number states
\citep{AaronsonArkhipov2013LV,broome2013photonic,crespi2013integrated,tillmann2013experimental,springBosonSamplingPhotonic2013a,spagnolo2014experimental,crespi2016suppression,Wang2019BosonSampling}
or Gaussian squeezed states \citep{Hamilton2017PhysRevLett.119.170501,kruse2019detailed,quesada2018gaussian,zhong2020quantum,zhongPhaseProgrammableGaussianBoson2021,madsenQuantumComputationalAdvantage2022}
to generate random, discrete counts by sampling matrix permanents,
Hafnians, or the Torontonian \citep{Aaronson2011,AaronsonArkhipov2013LV,Hamilton2017PhysRevLett.119.170501,quesada2018gaussian}.
Which distribution is sampled depends on the input state and detectors
used. All of these variants have output counting distributions which
are exponentially hard to either compute or sample for photonic networks
of large size \citep{Chabaud2021nonclassical,Chabaud2023Resources}.

In this paper, we expand upon earlier work which showed how one can
use grouped count probabilities (GCPs) to validate outputs of Gaussian
boson sampling (GBS) quantum computers with threshold detectors \citep{drummondSimulatingComplexNetworks2022,delliosSimulatingMacroscopicQuantum2022a}.
This is achieved with scalable simulations using phase-space mapping
to continuous samples. These techniques allow one to compare theoretical
and experimental output correlations and marginal probabilities, as
the simulations have identical moments and correlations to the ideal
GBS outputs. These methods have been scaled to exceptionally large
sizes of up to 16,000 modes \citep{drummondSimulatingComplexNetworks2022}.
Because these methods can include decoherence, they can model practical
experiments as well as ideal distributions. Here our quantum predictions
are compared with a 144-mode experiment claimed to display quantum
advantage. We explain how these techniques can distinguish quantum
data from classical fakes, even in the regime where there is decoherence
and noise present.

GBS experiments are in the hard domain for more than a hundred modes,
and are used for random number generation. There are other, much larger
bosonic networks in development. These are designed to solve hard
optimization problems with up to 100,000 modes, \citep{shoji2017quantum,wang2013coherent,yamamoto2017coherent,McMahon_CIM_science2019,yamamura2017quantum,Honjo2021Coherent},
or even large cluster states with up to a million modes \citep{yoshikawa2016invited}.
While such larger systems have more practical applications, the GBS
case has great scientific interest because quantum advantage is difficult
to prove. It has an architecture that allows a detailed theoretical
model. By comparing theory with experiment, one can understand how
to validate network-based quantum computers, and how to test for experimental
imperfections.

The successive large scale implementations of GBS quantum computers
claiming quantum supremacy \citep{zhong2020quantum,zhongPhaseProgrammableGaussianBoson2021,madsenQuantumComputationalAdvantage2022}
have lead to an outpacing of previous classical verification methods.
These either directly compute samples of output distributions such
as the Torontonian, for small mode numbers \citep{bulmerBoundaryQuantumAdvantage2022a,quesadaQuadraticSpeedUpSimulating2022},
or compute low-order marginal probabilities at larger mode numbers
\citep{villalonga2021efficient,ohSpoofingCrossEntropy2022a}. Practically,
these methods are applied to verify GBS outputs for small numbers
of modes, since neither method can verify all moments of experimental
networks. This is due to such classical methods encountering severe
computational barriers when computing high-order correlations, because
the full distribution itself is known to be a \#P-hard computational
problem. 

As network sizes continue to increase, verifying high-order correlations
becomes increasingly important. Despite this computational hardness,
a testing protocol is essential to ensure that experimental errors
such as parameter drift, decoherence and noise are negligible. Our
phase-space methods can validate all measurable correlations, because
they do not use discrete counts. Doing this would be the \#P-hard
computational problem implemented by the current generation of GBS
devices. The positive-P method overcomes this by generating continuous
random outputs with equivalent moments, with similar or better accuracy
to experimental measurement. Validating the outputs is a different
computational task from generating all discrete random counts.

Validation tests can also be used to show a classical imitation is
significantly different from the required output, to eliminate fakes.
To do this, our methods can generate exponentially many tests by randomly
permuting the output groups or bins. Such tests cannot all be implemented
at once. 

Any attempt to fake the output counts will encounter a computationally
hard ``shell-game''. The counterfeiter cannot predict which test
will be used. Thus, any classical algorithm designed just to deceive
a small number of such tests will fail, in all except an exponentially
small number of scenarios. While we cannot rigorously eliminate every
fake, we conjecture that these randomized tests are computationally
hard to pass using classical means, provided there is enough experimental
data.

Such tests were performed using GCPs that are binned in multiple dimensions.
Multi-dimensional comparisons allow both a fine tuned comparison of
experimental outputs with theory, and an exceptionally powerful method
to differentiate between fake classical algorithms and experimental
data. Comparisons are made with data from a 144-mode GBS experiment
using threshold detectors, with measurements of up to $133$-th order
correlations \citep{zhongPhaseProgrammableGaussianBoson2021}. To
demonstrate how grouped photon counts can differentiate between experimental
and faked correlations, we compare simulations of the ideal GBS output
with faking strategies that generate discrete random photon counts.
These are generated from classical squeezed thermal states \citep{Reid1986,Fearn_JModOpt1988},
also called squashed states \citep{jahangiriPointProcessesGaussian2020,martinez-cifuentesClassicalModelsAre2022},
which are input into the linear network. 

Comparisons of marginal click correlation moments are also presented,
as our numerical method allows one to efficiently generate comparisons
for \textit{all} possible output mode combinations. These are often
used to compare the accuracy of samples from experiments that claim
the presence of nontrivial correlations \citep{zhongPhaseProgrammableGaussianBoson2021},
with low-order marginal probability based classical algorithms \citep{villalonga2021efficient}. 

An analysis of scaling due to sampling errors generated from increased
correlation order is given, as normally-ordered positive-P phase-space
representations have no vacuum noise, and are therefore very efficient
in simulating photo-detection. Due to its reduced sampling errors,
the positive-P method \citep{Drummond_Gardiner_PositivePRep} is applicable
to existing large-scale data sets, and is exponentially faster than
non-normally ordered methods \citep{opanchuk2018simulating}. 
\begin{figure}
\begin{centering}
\includegraphics[height=1\columnwidth]{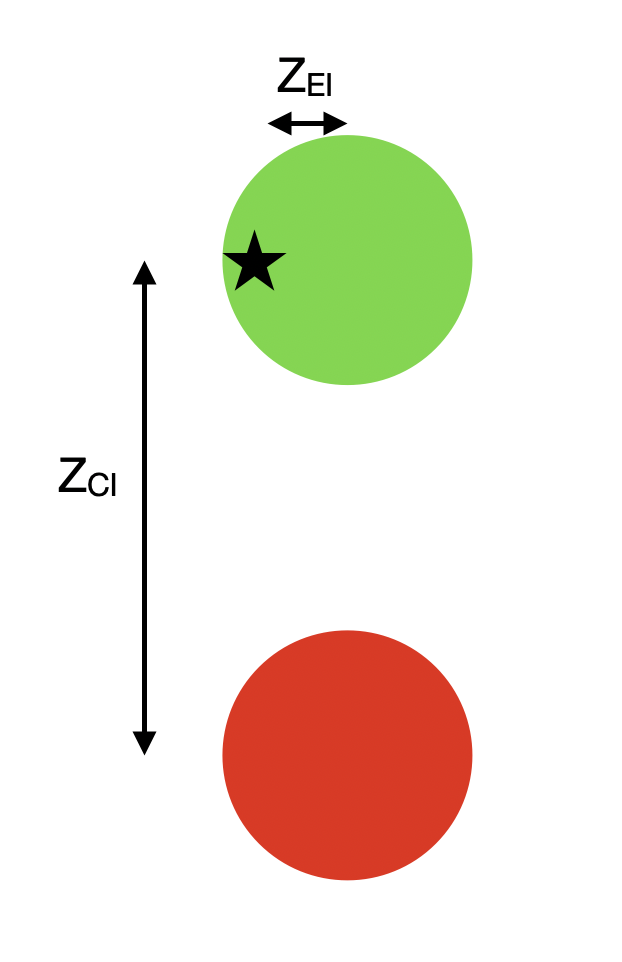}
\par\end{centering}
\caption{A schematic diagram of the goal of a GBS experiment, in which the
star indicates the ideal or target distribution (I), the green circle
the sampled distribution from an experiment, and the red circle the
nearest efficiently simulable distribution using classical methods.
Ideally, one should have $Z_{EI}\lesssim1$, while $Z_{CI}\gg1$.
This indicates that no classical computer can generate the required
distribution, while the quantum computer does. Here $Z$ is a statistical
distance measure obtained from $\chi^{2}$ test data, normalized primarily
by the experimental sampling errors. \label{fig:Ideal-picture}}
\end{figure}

The diagram of Figure (\ref{fig:Ideal-picture}) shows the goal of
the experiment, which is to demonstrate quantum computational advantage.
This requires that the experimental data agrees with the ideal theory,
$Z_{EI}\lesssim1$, while no efficient classical simulation can do
this, $Z_{CI}\gg1$. Here $Z$ measures the distribution distance
in units of the sampling error. Results of comparisons for all observables
using $\chi^{2}$ statistical tests demonstrate that the present 144-mode
experimental data set shows large deviations from the ideal GBS distribution,
with deviations of over $Z_{EI}\gg100$ for all data sets, indicating
the first requirement is not met. These deviations are highlighted
when classically generated binary patterns are compared to the ideal
distribution. These show better agreement than experiments with the
ideal model.

\begin{figure}
\begin{centering}
\includegraphics[height=1\columnwidth]{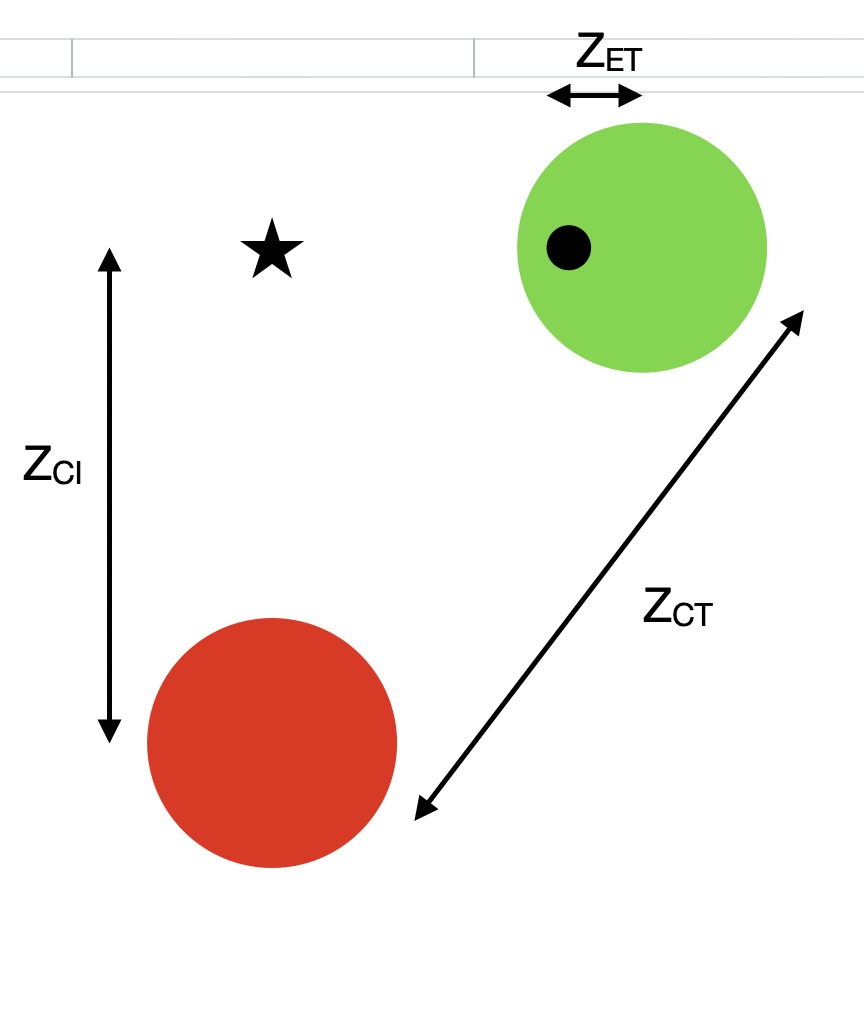}
\par\end{centering}
\caption{A schematic diagram of the ``best-case'' of current GBS experiments,
with red and green circles as in Fig (\ref{fig:Ideal-picture}), while
the black circle gives the thermalized distribution (T). In this realistic
case, one may require that $Z_{ET}\lesssim1$, while $Z_{CT}\gg1$.
This indicates that no known efficient classical algorithm can generate
the thermalized target distribution, while the quantum computer does.
\label{fig:Real-picture}}
\end{figure}

The comparison of theory to experiment is greatly improved when additional
thermalization is included in the squeezed input model. Hence, a more
realistic goal is to have $Z_{ET}\lesssim1$, with $Z_{CT}\gg1$,
which would indicate a computational advantage for the thermalized
model. While most of the data does not achieve this, it is obtained
for the total count distribution of one of the tested data sets. This
situation is indicated in Fig (\ref{fig:Real-picture}). 

\begin{figure}
\begin{centering}
\includegraphics[height=1\columnwidth]{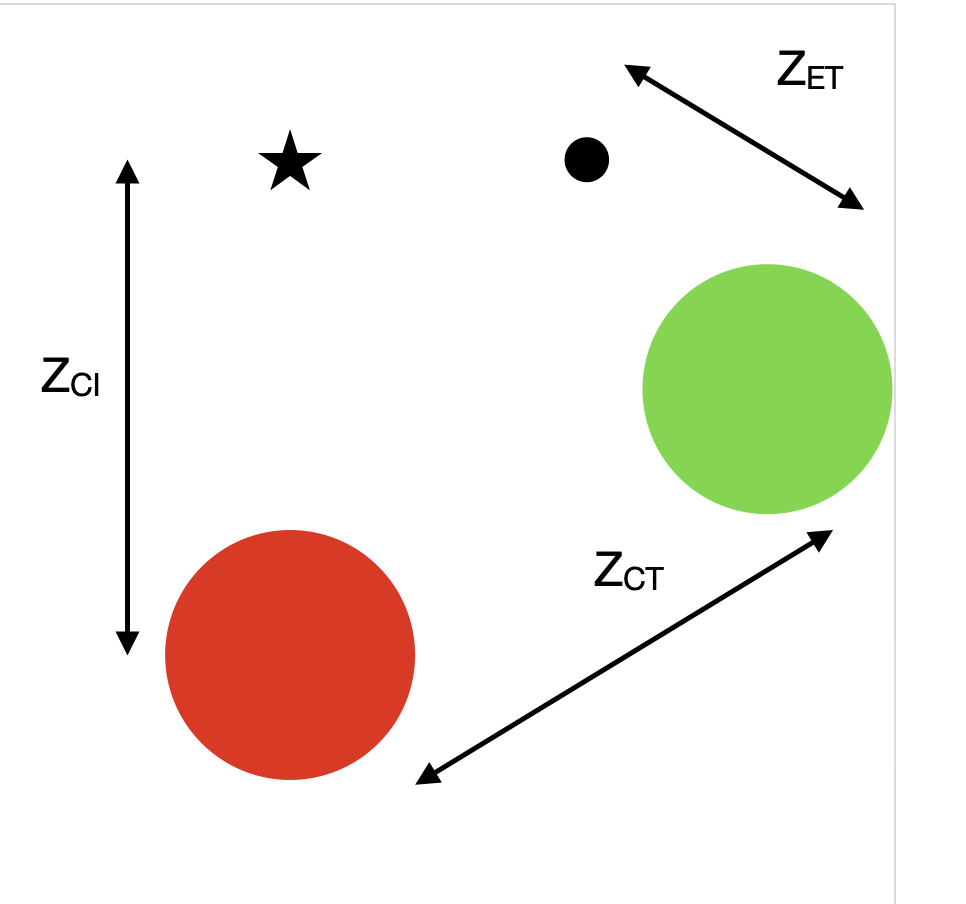}
\par\end{centering}
\caption{A schematic diagram of more typical results of current GBS experiments,
with color codes as in Fig (\ref{fig:Real-picture}). For this more
realistic case, one has $Z_{ET}>1$, giving a poor fit to a thermalized
model, but the classical model is even worse, such that $Z_{CT}\gg1$.
We emphasize that other, more efficient classical algorithms may exist,
which could alter these comparisons. \label{fig:Real-general-picture}}
\end{figure}

In other cases, while the residual differences are significant, when
classical photon counts are compared with this nonclassical but thermalized
model, their total count distributions are further from the thermalized
distribution than the experiment, shown schematically in Fig (\ref{fig:Real-general-picture}).

The residual differences may be caused by fluctuations of the network,
parameter estimation errors, or nonlinear effects, and are identified
by computing the $Z$-statistic. Comparisons of low-order marginals
are also performed. These are further from the ideal and thermalized
models than classical fakes for all tested data sets. However, high
order total count correlations appear to provide a more robust test
of GBS statistics than the low-order marginals, which are sensitive
to minor transmission parameter errors. Hence, such low-order correlation
differences are a poor test for quantum advantage.

An interesting open question in computer science, not treated here,
is whether all efficient (not exponentially slow) classical fakes
can be identified using the multiple statistical tests that we have
identified in this paper.

Our results highlight the importance of scalable validation methods
for experimental data in current technologies. These algorithms provide
techniques for validating large experiments. Scalability requirements
and computability are fundamental to any theory that describes high-order
multi-mode experiments. This includes Bose-Einstein condensates \citep{schweigler2017experimental},
dynamical phase-transitions \citep{Marino_2022}, noisy quantum computers
\citep{Blume-Kohout2022Taxonomy}, random quantum circuits \citep{takeuchi2018verification},
multi-qubit photon-atom interactions \citep{goldberg2022beyond},
coherent Ising machines (CIM) \citep{shoji2017quantum,Honjo2021Coherent},
and many others now under experimental development. 

In summary, grouped count probabilities simulated in phase-space can
be used to compare the experimental correlations of large-scale Gaussian
boson sampling experiments with quantum theoretical predictions. We
distinguish the experimental data from some types of classically generated
data. This is carried out using a statistical distance $Z$, normalized
by the sampling errors. High order correlations give the strongest
tests. There is substantial disagreement between experiment and the
idealized GBS model. However, evidence does exist for quantum advantage
from the high-order correlations, provided the target computational
distribution is slightly decoherent.

\section{Phase-space representations of bosonic networks}

We first summarize results presented previously \citep{drummondSimulatingComplexNetworks2022,delliosSimulatingMacroscopicQuantum2022a}
on representing the input and output states of a bosonic network with
phase-space methods. Such representations are a natural fit for describing
bosonic networks with Gaussian inputs. They are inherently scalable
and have analytical expressions which are simple to implement numerically. 

They are also applicable to other quantum technologies with nonlinearities
and feedback, like the CIM \citep{maruo2016truncated,shoji2017quantum,kiesewetter2022phase,Kiesewetter2022Coherent}.
To simulate quantum inputs, we focus on the generalized P-representation
\citep{Drummond_generalizedP1980}, Wigner representation \citep{Wigner_1932,Moyal_1949}
and Q-function \citep{Husimi1940} which can all give positive, non-singular
distributions for squeezed state inputs. 

Choosing a representation that minimizes computational sampling errors
is of paramount importance. We will show that the normally ordered
positive-P method is the preferred choice for GBS photon-counting
experiments, due to its low sampling errors for high-order correlations,
as shown in Section (\ref{subsec:intensity correlations}). 

\subsection{Input state\label{subsec:Input-state}}

Linear networks are conceptually very simple. Without losses, the
network itself is represented by a $M\times M$ Haar random unitary
matrix $\boldsymbol{U}$ \citep{AaronsonArkhipov2013LV,Hamilton2017PhysRevLett.119.170501,kruse2019detailed,hangleiterComputationalAdvantageQuantum2022},
however losses cause the network to become non-unitary. Therefore,
a lossy network is denoted by the $M\times M$ transmission matrix
$\boldsymbol{T}$. Out of $M$ total input channels, $N\subset M$
are filled with input states, which are then converted to $M$ outputs
via the linear network. 

\subsubsection{Pure squeezed states}

In an ideal GBS experiment, the inputs are independent Gaussian single-mode
squeezed states, allowing one to write the input state as $\hat{\rho}^{(\text{in})}=\prod_{j}\left|\boldsymbol{r}\right\rangle \left\langle \boldsymbol{r}\right|$,
where $\boldsymbol{r}=\left[r_{1},\dots,r_{N}\right]$ is the squeezing
vector. Ideally, these inputs are pure squeezed states, which are
nonclassical minimum uncertainty states defined entirely by their
quadrature variances \citep{yuen1976two,Drummond2004_book,vahlbruch2016detection}. 

Following standard quantum optics techniques \citep{drummond2014quantum},
the non-vanishing quadrature correlations are 
\begin{align}
\left\langle \left(\Delta\hat{x}_{j}\right)^{2}\right\rangle  & =2\left(n_{j}+m_{j}\right)+1=e^{2r_{j}}\nonumber \\
\left\langle \left(\Delta\hat{y}_{j}\right)^{2}\right\rangle  & =2\left(n_{j}-m_{j}\right)+1=e^{-2r_{j}}.\label{eq:pure squeeze variance}
\end{align}
Here, $\hat{x}_{j}=\hat{a}_{j}+\hat{a}_{j}^{\dagger}$, $\hat{y}_{j}=\left(\hat{a}_{j}-\hat{a}_{j}^{\dagger}\right)/i$
are the quadrature operators which obey the commutation relation $\left[\hat{x}_{j},\hat{y}_{k}\right]=2i\delta_{jk}$,
while $n_{j}=\sinh^{2}(r_{j})$ and $m_{j}=\cosh(r_{j})\sinh(r_{j})$
are the input photon number and coherence per mode, respectively. 

From the Heisenberg uncertainty principle, this allows one to write
the requirement for a minimum uncertainty ideal squeezed state as
\begin{equation}
\left\langle \left(\Delta\hat{x}_{j}\right)^{2}\right\rangle \left\langle \left(\Delta\hat{y}_{j}\right)^{2}\right\rangle =1.
\end{equation}

\subsubsection{Thermalized squeezed states }

Experimentally generating pure squeezed states is challenging. Laboratory
equipment such as lasers, polarizing beamsplitters, mirrors and phase-shifters
will inevitably introduce decoherence due to laser noise, temporal
drift, refractive index fluctuations \citep{Perlmutter1990PhysRevB.42.5294},
mode mismatch \citep{zhong2020quantum} and dephasing effects \citep{drummond2020initial}. 

This means that the squeezed states can no longer be considered pure,
and realistically one has $\left\langle \left(\Delta\hat{x}_{j}\right)^{2}\right\rangle \left\langle \left(\Delta\hat{y}_{j}\right)^{2}\right\rangle >1$.
Therefore, to accurately model an experimental implementation of bosonic
networks, one needs to account for this additional decoherence, even
though it is generally not included in the reported experimental data.
We do this by fitting the reported data to a model for thermalized
squeezed states \citep{Fearn_JModOpt1988}, together with a correction
to the transmission matrix \citep{drummondSimulatingComplexNetworks2022}.

We suppose that a beamsplitter attenuates the input intensity by a
factor of $1-\epsilon$, while adding $n_{j}^{th}=\epsilon n(r_{j})$
thermal photons per mode. This alters the input coherence as $\tilde{m}_{j}=\left(1-\epsilon\right)m(r_{j})$,
whilst keeping the input photon number unchanged. The advantage of
this model is that one can easily test a variety of input states from
thermal, $\epsilon=1$, to pure squeezed states, $\epsilon=0$, and
anything in between, by simply changing $\epsilon$. Because this
changes the resulting count distribution, we improve the fit to the
reported data by including a correction factor $t$ to the transmission
matrix.

Since squeezed states can be modeled using a variety of phase-space
methods, it is useful to employ operator-ordering methods for such
simulations. The most convenient method uses the amount of vacuum
noise added with each representation to define a corresponding operator
ordering parameter $\sigma$, which is similar to $s$-ordering \citep{Cahill_Glauber_OrderedExpansion_1969}.
Here, $\sigma=0$ denotes normal ordering, $\sigma=1/2$ symmetric
ordering and $\sigma=1$ anti-normal ordering. 

Using this ordering method, the squeezed quadrature variance with
the above beamsplitter model of decoherence in any representation
is defined as 

\begin{align}
\left\langle \left\{ \left(\Delta\hat{x}_{j}\right)^{2}\right\} _{\sigma}\right\rangle  & =\Delta_{\sigma x_{j}}^{2}=2\left(n_{j}+\sigma+\tilde{m}_{j}\right)\nonumber \\
\left\langle \left\{ \left(\Delta\hat{y}_{j}\right)^{2}\right\} _{\sigma}\right\rangle  & =\Delta_{\sigma y_{j}}^{2}=2\left(n_{j}+\sigma-\tilde{m}_{j}\right).\label{eq:sigma variance}
\end{align}

For compact notation, we may omit the $\sigma$ in subscripts when
it is zero, for normal ordering.

\subsection{Glauber-Sudarshan P-representation}

To simulate linear networks in phase-space, one is restricted by both
the input state and the type of detector used. If normally ordered
photo-number-resolving (PNR) detectors are used, any non-normally
ordered representation introduces vacuum noise in the initial stochastic
samples. We show in Sec. (\ref{subsec:intensity correlations}) that
this causes a rapid growth of computational sampling errors when computing
high-order intensity correlations.

We first summarize results for the diagonal Glauber-Sudarshan P-representation,
which is defined in terms of the density matrix \citep{Glauber_1963_P-Rep,Sudarshan_1963_P-Rep}
using coherent states $\left|\boldsymbol{\alpha}\right\rangle $,
as:

\begin{equation}
\hat{\rho}=\int P(\boldsymbol{\alpha})\left|\boldsymbol{\alpha}\right\rangle \left\langle \boldsymbol{\alpha}\right|\text{d}^{2}\boldsymbol{\alpha}.
\end{equation}
This normally-ordered phase-space representation can have singular
distributions for general quantum states, including squeezed and number
states. However, it is always positive for thermal and coherent states
\citep{Drummond2004_book,walls2008quantum}. More generally, classical
states are defined as having a positive diagonal P-representation,
so that no quadrature has a variance below that of the vacuum state
\citep{Reid1986,RahimiKeshari:2016}. 

\subsubsection{Classical states}

Because they generate discrete counts that are classically stimulable,
classical states have been analyzed to verify experiments indeed send
$N$ quantum states into the linear network. Large amounts of decoherence
in the inputs may cause the input state to become classical, allowing
the resulting output distribution to be efficiently simulated on a
classical computer \citep{RahimiKeshari:2016,qi2020regimes}. 

An extreme case is a thermal state, which is a fully decoherent classical
state with normally-ordered quadrature variance

\begin{equation}
\left\langle \left(\Delta\hat{x}_{j}\right)^{2}\right\rangle =\left\langle \left(\Delta\hat{y}_{j}\right)^{2}\right\rangle =2n_{j},\label{eq:thermal_variance}
\end{equation}
obtained by letting $\epsilon=1$ in Eq.(\ref{eq:sigma variance}). 

Simulations of $N\subset M$ thermal states sent into a linear network
have been performed previously \citep{zhong2020quantum,zhongPhaseProgrammableGaussianBoson2021}
and shown to not accurately model any recent experimental implementations
of GBS \citep{zhong2020quantum,zhongPhaseProgrammableGaussianBoson2021,madsenQuantumComputationalAdvantage2022}. 

A more realistic state is a classical approximation to pure squeezed
states called squashed states \citep{jahangiriPointProcessesGaussian2020,martinez-cifuentesClassicalModelsAre2022}.
These states arise as the classical limit of thermalized squeezed
states with $\epsilon=\lim_{r_{j}\rightarrow\infty}(1-\tanh(r_{j}))$,
corresponding to $m_{j}=n_{j}$. Unlike thermal states, squashed states
maintain the squeezed quadrature variance condition 

\begin{equation}
\left\langle \left(\Delta\hat{x}_{j}\right)^{2}\right\rangle \neq\left\langle \left(\Delta\hat{y}_{j}\right)^{2}\right\rangle ,
\end{equation}
however, neither quadrature is squeezed below the vacuum noise limit.
From Eq.(\ref{eq:pure squeeze variance}) the normally ordered variance
is defined as \citep{jahangiriPointProcessesGaussian2020,martinez-cifuentesClassicalModelsAre2022}:

\begin{align}
\left\langle \left(\Delta\hat{x}_{j}\right)^{2}\right\rangle  & =4n_{j}+1\nonumber \\
\left\langle \left(\Delta\hat{y}_{j}\right)^{2}\right\rangle  & =1.\label{eq:squashed_variance}
\end{align}

Although squashed states contain no true squeezing as one quadrature
has fluctuations at the vacuum limit, squashed states present a more
realistic classical input state compared to the fully decoherent thermal
states. 

Recently, simulations of squashed states input to the $100$-mode
bosonic network of Zhong et al \citep{zhong2020quantum} were shown
to be closer to the experimental output distributions than the theoretical
ideal GBS distribution \citep{martinez-cifuentesClassicalModelsAre2022}.
However the same simulations performed for the $144$-mode network
of Zhong et al \citep{zhong2020quantum} produced mixed results, as
outlined in more detail in section \ref{sec:Classical_fake}. 

Although the diagonal P-representation is unsuitable for simulating
networks with squeezed state inputs, it is well suited to simulate
networks with classical inputs \citep{qi2020regimes,RahimiKeshari:2016}.
This is demonstrated in section \ref{sec:Classical_fake} for squashed
states which are also used to generate fake binary patterns. 

The detailed form of the distribution is given in the next subsection.

\subsection{Wigner and Q representations}

For $\sigma\ge1/2$ one finds that a classical phase-space is always
sufficient to obtain a non-negative Gaussian distribution, even for
squeezed states. This leads to the symmetrically ordered Wigner representation
($\sigma=1/2$) and anti-normally ordered Q-function ($\sigma=1$),
which are other alternatives. Both are defined on a classical phase-space
and generate a positive distribution for any Gaussian input state. 

For Gaussian states, the Wigner distribution can be written in the
simple form \citep{Louisell,Hillery_Review_1984_DistributionFunctions,drummond2014quantum}

\begin{equation}
W\left(\boldsymbol{\alpha}\right)=\frac{1}{\pi^{2N}}\int\text{d}^{2}\boldsymbol{\mathrm{z}}\text{Tr}\left\{ \hat{\rho}e^{i\boldsymbol{\mathrm{z}}\cdot\left(\hat{\boldsymbol{\mathrm{a}}}-\boldsymbol{\alpha}\right)+i\boldsymbol{\mathrm{z}}^{*}\cdot\left(\hat{\boldsymbol{\mathrm{a}}}^{\dagger}-\boldsymbol{\alpha}^{*}\right)}\right\} ,\label{eq:Wigner distribution}
\end{equation}
where $\text{Tr}\{\dots\}$ is a trace and $\boldsymbol{\mathrm{z}}$
is a complex vector, while the Q-function is written in the standard
form \citep{Husimi1940}: 

\begin{equation}
Q\left(\boldsymbol{\alpha}\right)=\frac{1}{\pi^{N}}\left\langle \boldsymbol{\alpha}\mid\hat{\rho}\mid\boldsymbol{\alpha}\right\rangle .
\end{equation}

These representations have been used previously to obtain analytical
expressions for the probability of a specific GBS output pattern \citep{Hamilton2017PhysRevLett.119.170501,kruse2019detailed,quesada2018gaussian}
and to determine the classical simulability of noisy GBS networks
\citep{qi2020regimes}. We use the notation $P_{\sigma}$ to indicate
any distribution of this type over a classical phase-space.

In all cases, we define classical quadrature phase-space variables
as:
\begin{align}
x_{j} & =\alpha_{j}+\alpha_{j}^{*}\sim\hat{x}_{j},\nonumber \\
y_{j} & =\left(\alpha_{j}-\alpha_{j}^{*}\right)/i\sim\hat{y}_{j}.
\end{align}

Any thermalized squeezed vacuum state with non-negative variances
$\Delta_{\sigma x_{j}}^{2}$ , $\Delta_{\sigma y_{j}}^{2}$ has a
non-singular, Gaussian distribution on a classical phase-space, with
\begin{equation}
P_{\sigma}(\boldsymbol{\alpha})=\prod_{j}\left(\frac{2}{\pi\Delta_{\sigma x_{j}}\Delta_{\sigma y_{j}}}e^{-x_{j}^{2}/2\Delta_{\sigma x_{j}}^{2}-y_{j}^{2}/2\Delta_{\sigma y_{j}}^{2}}\right).\label{eq:sigma_DP_distribution}
\end{equation}

Using the $\sigma$-ordering scheme, the equivalence between operator
moments and stochastic moments is given by the $\sigma$-ordering
relation:
\begin{align}
\left\langle \left\{ \hat{a}_{j_{1}}^{\dagger},\ldots,\hat{a}_{j_{n}}\right\} _{\sigma}\right\rangle  & =\left\langle \alpha_{j_{1}}^{*},\ldots,\alpha_{j_{n}}\right\rangle _{\sigma}\nonumber \\
 & =\int P_{\sigma}(\boldsymbol{\alpha})\left[\alpha_{j_{1}}^{*},\ldots,\alpha_{j_{n}}\right]\text{d}^{2M}\bm{\alpha}.\label{eq:sigma moment equivalence}
\end{align}
where $\left\{ \dots\right\} _{\sigma}$ and $\left\langle \dots\right\rangle _{\sigma}$
denotes $\sigma$-ordered operator products and stochastic averages,
respectively. 

Representations with $\sigma\ge1/2$ introduce vacuum noise in the
initial stochastic samples when used to analyze photon-number detectors.
The additional noise makes the Wigner and Q representations completely
impractical for any computation of high-order correlations in current
large-scale bosonic networks that use photon-number detectors. 

We show below that the added vacuum noise causes an exponential increase
in sampling error with $M$, for high-order correlations.

\subsection{Positive P-representation}

The generalized P-representation is a normally ordered distribution
in phase-space that is exact and non-singular for any input quantum
state. This is useful for simulating the correlations of squeezed
or number states, as it doesn't introduce vacuum noise. 

The representation is written as 

\begin{equation}
\hat{\rho}=\int\int P\left(\boldsymbol{\alpha},\boldsymbol{\beta}\right)\hat{\Lambda}\left(\boldsymbol{\alpha},\boldsymbol{\beta}\right)\text{d}\mu\left(\boldsymbol{\alpha},\boldsymbol{\beta}\right),\label{eq:gen P-rep}
\end{equation}
where $\hat{\rho}$ is expanded over a subspace of the complex plane,
$\boldsymbol{\alpha},\boldsymbol{\beta}$ are independent coherent
state amplitude vectors \citep{Glauber1963_CoherentStates} and

\begin{equation}
\hat{\Lambda}\left(\boldsymbol{\alpha},\boldsymbol{\beta}\right)=\frac{\left|\boldsymbol{\alpha}\right\rangle \left\langle \boldsymbol{\beta}^{*}\right|}{\left\langle \boldsymbol{\beta}^{*}\mid\boldsymbol{\alpha}\right\rangle }
\end{equation}
is the off-diagonal coherent state projector. If $P\left(\boldsymbol{\alpha},\boldsymbol{\beta}\right)=P\left(\boldsymbol{\alpha}\right)\delta\left(\boldsymbol{\alpha}^{*}-\boldsymbol{\beta}\right)$,
this reduces to the Glauber-Sudarshan representation where $\boldsymbol{\beta}=\boldsymbol{\alpha}^{*}$
defines a classical phase-space.

The projection operator projects the density matrix onto multi-mode
coherent states. This is responsible for the exact and non-singular
nature of the generalized-P distribution for quantum inputs as it
doubles the classical phase-space dimension, which allows off-diagonal
coherent state amplitudes with $\boldsymbol{\beta}\neq\boldsymbol{\alpha}^{*}$
to exist in the basis. These represent nonclassical quantum superposition
states \citep{Drummond_generalizedP1980,Drummond:2016}. 

The generalized P-representation is a family of normally ordered representations
with different distributions $P\left(\boldsymbol{\alpha},\boldsymbol{\beta}\right)$,
the form of which is dependent on the integration measure $\text{d}\mu\left(\boldsymbol{\alpha},\boldsymbol{\beta}\right)$
\citep{Drummond_generalizedP1980}. Here, we use the positive P-representation,
which is obtained when $\text{d}\mu\left(\boldsymbol{\alpha},\boldsymbol{\beta}\right)=\text{d}^{2M}\boldsymbol{\alpha}\text{d}^{2M}\boldsymbol{\beta}$,
which is a $4M$-dimensional volume integral, and $\boldsymbol{\alpha},\boldsymbol{\beta}$
can vary along the whole complex plane. By taking the real part of
Eq.(\ref{eq:gen P-rep}), the density matrix becomes hermitian and
can be sampled efficiently. 

Because it gives an efficiently sampled, non-singular and strictly
positive output distribution in all cases, the positive P-representation
is ideal for simulating bosonic networks with squeezed state inputs.
It combines probabilistic properties with operator normal-ordering,
giving a one-to-one relationship between normally-ordered operator
moments and stochastic moments \citep{Drummond_Gardiner_PositivePRep}:

\begin{equation}
\left\langle \hat{a}_{j_{1}}^{\dagger},\ldots,\hat{a}_{j_{n}}\right\rangle =\left\langle \beta_{j_{1}},\ldots,\alpha_{j_{n}}\right\rangle _{P}.\label{eq:+P op equiv}
\end{equation}
This relationship is valid for any generalized P-representation, where
$\left\langle \dots\right\rangle $ denotes a quantum expectation
value and $\left\langle \dots\right\rangle _{P}$ is a generalized-P
average. 

For a pure squeezed state, the input state density matrix $\hat{\rho}^{(\text{in})}$
can be written in terms of the positive-P distribution by expanding
each squeezed state $\left|\boldsymbol{r}\right\rangle $ as a line
integral over a real coherent state \citep{adam1994complete}, so
that $\text{d}\mu\left(\boldsymbol{\alpha},\boldsymbol{\beta}\right)=\text{d}\boldsymbol{\alpha}\text{d}\boldsymbol{\beta}$,
with $\boldsymbol{\alpha},\boldsymbol{\beta}$ as independent real
vectors. Alternatively, one can view this as a distribution over the
$4M$ dimensions of the full phase-space, with delta-function distributions
on the $2M$ imaginary axes.

This gives 
\begin{equation}
\hat{\rho}^{(\text{in})}=\text{Re}\int\int P\left(\boldsymbol{\alpha},\boldsymbol{\beta}\right)\hat{\Lambda}\left(\boldsymbol{\alpha},\boldsymbol{\beta}\right)\text{d}\boldsymbol{\alpha}\text{d}\boldsymbol{\beta}.
\end{equation}
Here 
\begin{equation}
P\left(\boldsymbol{\alpha},\boldsymbol{\beta}\right)=\prod_{j}C_{j}e^{-\left(\alpha_{j}^{2}+\beta_{j}^{2}\right)\left(\gamma^{-1}+1/2\right)+\alpha_{j}\beta_{j}}\label{eq:squeezed positive P}
\end{equation}
is a positive-P distribution for an input pure squeezed state, which
is a Gaussian distribution on a positive $\alpha,\beta$ plane, where
$C_{j}$ is the normalization constant and $\gamma_{j}=e^{2r_{j}}-1=2\left(n_{j}+m_{j}\right)$.
To diagonalize this Gaussian, we take $x_{j}=\alpha_{j}+\beta_{j}\sim\hat{x}_{j}$,
$\bar{y}_{j}=\alpha_{j}-\beta_{j}\sim i\hat{y}_{j}$ as real variables,
with the result that each is now an independent Gaussian, with
\begin{align}
\Delta_{x_{j}}^{2} & =2\left(n_{j}+m_{j}\right)\nonumber \\
\Delta_{\bar{y}_{j}}^{2} & =2\left(m_{j}-n_{j}\right).
\end{align}

This variable change gives the expansion as:
\begin{equation}
P\left(\boldsymbol{\alpha},\boldsymbol{\beta}\right)=\prod_{j}\left(\frac{2}{\pi\Delta_{x_{j}}\Delta_{\bar{y}_{j}}}e^{-x_{j}^{2}/2\Delta_{x_{j}}^{2}-\bar{y}_{j}^{2}/2\Delta_{\bar{y}_{j}}^{2}}\right).
\end{equation}

So far, we have assumed the squeezing orientation $\left\langle :\left(\Delta\hat{x}_{j}\right)^{2}:\right\rangle >0$
and $\left\langle :\left(\Delta\hat{y}_{j}\right)^{2}:\right\rangle <0$,
as it is for a pure squeezed state. Since one of the normally-ordered
variances is always negative, the phase-space variable corresponding
to the hermitian $\hat{y}$ operator is imaginary, which requires
that $\alpha^{*}\neq\beta$. Hence, we have a nonclassical phase-space.

This result can be extended to thermalized cases by modifying the
variances, as long as $\left\langle :\left(\Delta\hat{y}_{j}\right)^{2}:\right\rangle <0$.
If thermalization is stronger, with $\left\langle :\left(\Delta\hat{y}_{j}\right)^{2}:\right\rangle \ge0$,
then the integration domain is changed so that $\beta=\alpha^{*}$.
This reduces to the Glauber-Sudarshan classical case already treated.

\subsection{Gaussian sampling in $\sigma$-ordered representations\label{subsec:initial samples}}

The above results can be combined to give a unified random sampling
expression valid in the case of any Gaussian input. We can construct
initial stochastic samples, which are valid for any ordering $\sigma$,
as \citep{drummond2020initial}

\begin{align}
\alpha_{j} & =\frac{1}{2}\left(\Delta_{\sigma x_{j}}w_{j}+i\Delta_{\sigma y_{j}}w_{j+M}\right)\nonumber \\
\beta_{j} & =\frac{1}{2}\left(\Delta_{\sigma x_{j}}w_{j}-i\Delta_{\sigma y_{j}}w_{j+M}\right),\label{eq:sigma_input_stochastic}
\end{align}
where $\left\langle w_{j}w_{k}\right\rangle =\delta_{jk}$ are real
Gaussian noises. For a squeezed $y$-quadrature with normal ordering
where $\Delta_{y_{j}}$ is imaginary, $\alpha_{j}$ and $\beta_{j}$
are real and independent. This holds even for impure states. For cases
where $\Delta_{\sigma y_{j}}$is real, either because of thermalization
or because the ordering has $\sigma\ge1/2$, $\alpha_{j}$ and $\beta_{j}$
are complex conjugate.

This sampling method is able to generate any Gaussian state with no
cross-correlations between the $\hat{x}$ and $\hat{y}$ quadratures,
which are generically thermalized squeezed states. If there is no
squeezing below the vacuum level, this representation reduces to the
classical-like Glauber P-representation for normal ordering.

It is possible that even more sophisticated models are needed to fully
explain the current experimental observations, as explained below,
but that is outside the scope of the present paper.

\subsection{Output density matrix\label{subsec:Output-density-matrix}}

Practically, linear networks consist of a series of polarizing beamsplitters
and mirrors, causing the $N$ input modes to interfere, generating
large amounts of entangled states, and converting the input state
to the output state $\hat{\rho}^{(\text{out})}$. 

In terms of phase-space distributions, this corresponds to transforming
the initial stochastic amplitudes as $\boldsymbol{\alpha}'=\boldsymbol{T}\boldsymbol{\alpha}$
and $\boldsymbol{\beta}'=\boldsymbol{T}^{*}\boldsymbol{\beta}$, which
is valid for all representations, provided there are no losses. In
the normally ordered case, the resulting output density matrix for
nonclassical inputs can therefore be sampled as before, but with a
transformed projector:
\begin{equation}
\hat{\rho}^{(\mathrm{out})}=\text{Re}\int\int P(\bm{\alpha},\bm{\beta})\hat{\Lambda}\left(\bm{T}\bm{\alpha},\bm{T}^{*}\bm{\beta}\right)\mathrm{d}\mu\left(\bm{\alpha},\bm{\beta}\right).\label{eq:output P}
\end{equation}

To take into account losses and detector inefficiencies, one can include
a larger unitary with loss channels, but only consider the sub-matrix
of $\bm{T}$ for the channels that are measured. For example, in the
matrix $\bm{T}=t\bm{U}$, all channels experience equal loss where
$t$ is an amplitude transmission coefficient. Due to the normal
ordering property of the any P-representation, this method remains
exactly equivalent to using a master equation method to treat losses.
However, non-normally ordered methods require extra noise terms if
there are losses \citep{delliosSimulatingMacroscopicQuantum2022a,drummondSimulatingComplexNetworks2022}.

Thermal noise or other random processes can also be included if present.
For cases in which $n^{th}>0$ in the loss reservoirs or $\sigma>0$
one must include these additional quantum or thermal noise terms with
losses. Such terms correspond to $\sigma$-ordered noise in the reservoir
modes. For the results given here, we assume that thermal noise only
occurs in the input modes, and that the reservoirs are at zero temperature,
which is a good approximation in optical experiments. 

\section{Grouped Correlation probabilities\label{sec:Grouped-Correlation-probabilitie}}

Correlations provide a signature of measurable quantum states. For
these to be a useful signature, they must be readily observable, relevant
to interesting quantum features, and have a low enough sampling error
to provide an unambiguous result. In this section, we review both
Glauber intensity correlations and GCPs of bosonic networks, which
have already been successfully used to compare theory and experiment
for an $M=100$ mode GBS experiment \citep{drummondSimulatingComplexNetworks2022}.

\subsection{Intensity correlations\label{subsec:intensity correlations}}

The most commonly used correlation in quantum optics is the $n$-th
order Glauber intensity correlation \citep{Glauber1963_CoherentStates}.
In photonic experiments with PNR detectors \citep{Tichy2014,Mayer2011,Walschaers2016_NJP18,opanchuk2018simulating},
the expectation value of the product of normally ordered output number
operators in a set of up to $M$ output modes is observed:

\begin{equation}
G^{(n)}(c_{j})=\left\langle :(\hat{n}'_{j})^{c_{j}}\dots\:(\hat{n}'_{M})^{c_{M}}:\right\rangle ,
\end{equation}
where $\hat{n}'_{j}=\hat{a}_{j}^{\dagger\text{(out)}}\hat{a}_{j}^{\text{(out)}}$
is the output photon number operator, while $c_{j}=0,1,2,\dots$ is
the number of photon counts at the $j$-th detector, and $n=\sum c_{j}$
is the correlation order. 

In the positive-P phase-space representation, output correlations
are obtained by computing moments which, due to the equivalence of
operator and stochastic moments, are obtained simply by replacing
$\hat{n}'_{j}$ with $n'_{j}=\alpha'_{j}\beta'_{j}$, such that for
a large number of samples
\begin{equation}
G^{(n)}=\left\langle (n'_{j})^{c_{j}}\dots\:(n'_{M})^{c_{M}}\right\rangle _{0}.
\end{equation}

In the $\sigma$-ordered phase-space case, the required reordering
of all number operators produces a correction term which must be included
to remove the vacuum noise introduced by each operator. Provided $c_{j}=0,1$,
this correction allows the stochastic variable to become equivalent
to the normally ordered output particle number, when $\boldsymbol{T}$
is unitary, via the replacement

\begin{equation}
n'_{j}=\alpha'_{j}\beta'_{j}-\sigma.
\end{equation}

In principle, $c_{j}$ is arbitrary but is limited to $c_{j}=0,1$
for simple cross-correlations of photon-number resolved detectors.
For more general cases of higher order moments and correlations with
$c_{j}>1$ the non-normally-ordered expressions become cumbersome,
and are not listed here.

\begin{figure}
\begin{centering}
\includegraphics[width=1\columnwidth]{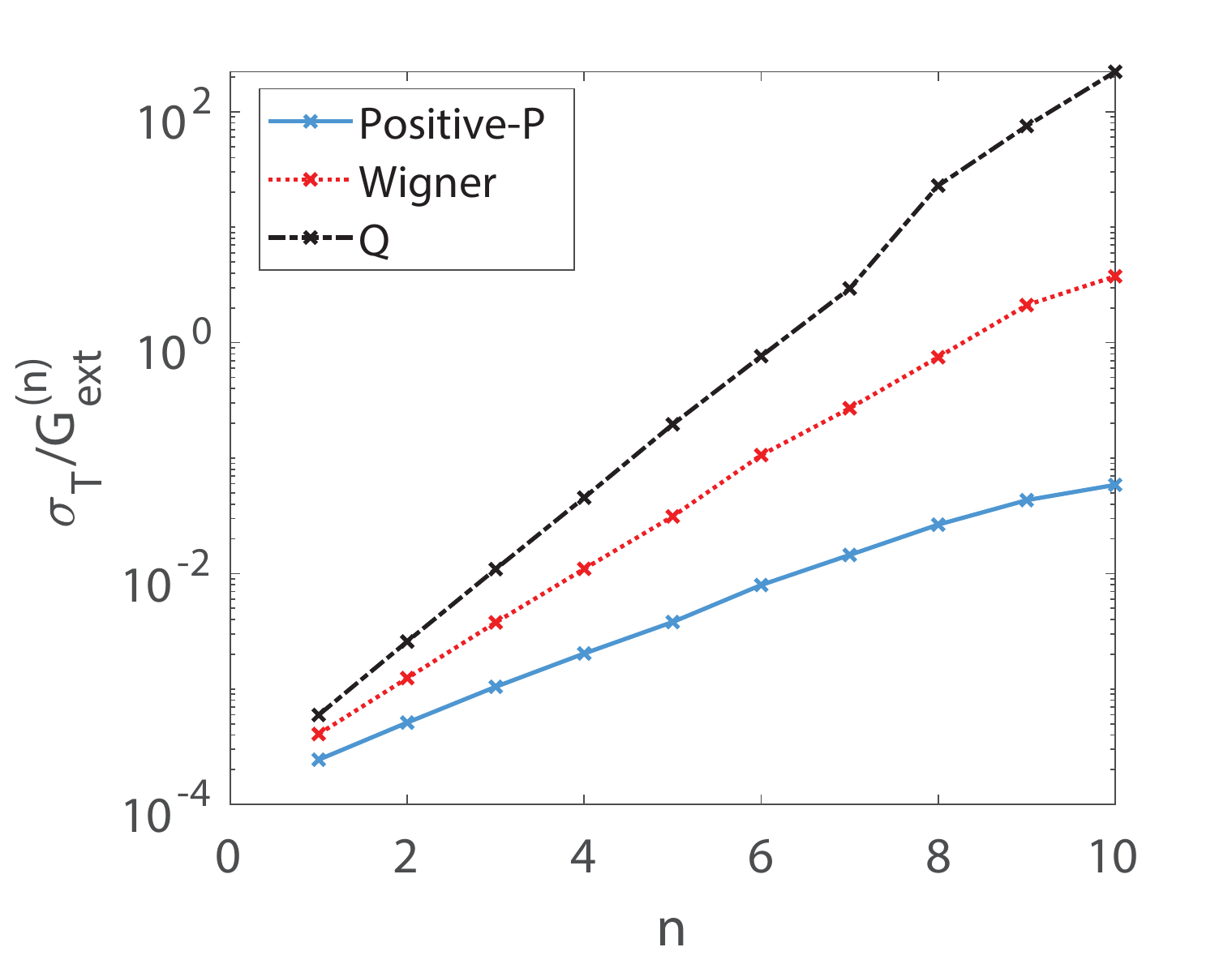}
\par\end{centering}
\caption{Comparison of sampling error growth for an $M=20$, $N=20$ GBS with
uniform pure squeezed state inputs and $E_{S}=4.8\times10^{7}$ samples.
The ratio of theoretical sampling errors, $\sigma_{T}$, with an exactly
computed n-th order intensity correlation, $G_{ext}^{(n)}$, is plotted.
Q-function simulations are denoted by the dashed black line, and add
the largest amount of vacuum noise per photon number. The Wigner representation
simulations are denoted by the red dotted line, which adds half a
quantum of noise per photon number, whilst the positive-P representation
adds no vacuum noise and corresponds to the solid blue line, with
far lower sampling error. Details are given in Section (\ref{sec:Sampling-errors-=000026}).
\label{fig:sampling error growth}}
\end{figure}

This in itself may not be a severe limitation, as GBS proposals with
PNR detectors often assume the probability of observing more than
one photon at a detector is small \citep{Hamilton2017PhysRevLett.119.170501,kruse2019detailed}.
However, as shown in Fig (\ref{fig:sampling error growth}), the computational
sampling error of Wigner and Q-function simulations grows rapidly
with correlation order, making them unsuitable for generating moments
to compare with experiment. 

For cases with sufficiently low flux corresponding to small mean photon
numbers, threshold detectors are equivalent to PNR detectors. Therefore,
intensity correlations are also the probability of an $N$-fold coincidence
count $P_{N}$. This allows one to write the correlation as a simple
product of output number operators, such that 

\begin{equation}
P_{N}\equiv\left\langle \prod_{j}\hat{n}'_{j}\right\rangle .
\end{equation}
At high flux levels, a single PNR detector may register more than
one count and the output is no longer binary. In such cases, we must
distinguish between PNR and threshold detectors. To get accurate results
for threshold detectors, a different operator is used.

\subsection{Grouped count correlations for saturating detectors}

To date, multiple GBS experiments of large scale networks have been
conducted using both PNR detectors \citep{madsenQuantumComputationalAdvantage2022},
and threshold, or click, detectors which saturate for more than one
count at a detector \citep{zhong2020quantum,zhongPhaseProgrammableGaussianBoson2021}.
When PNR detectors are used, one samples from the Hafnian distribution
\citep{Hamilton2017PhysRevLett.119.170501,kruse2019detailed}, whilst
threshold detectors are equivalent to sampling from the Torontonian
distribution \citep{quesada2018gaussian}. 

We focus on experiments using the latter detector type, with outputs
being binary numbers where the $j$-th detector records $c_{j}=1$
for a photon detection event, or click, and $c_{j}=0$ for no detection
event. Therefore, a network of $M$ detectors will produce binary
patterns represented by the count vector $\boldsymbol{c}$, with $2^{M}$
possible patterns available. Each detector is defined by the normally-ordered
projection operator \citep{Sperling2012True}

\begin{equation}
\hat{\pi}_{j}\left(c_{j}\right)=:e^{-\hat{n}'_{j}}\left(e^{\hat{n}'_{j}}-1\right)^{c_{j}}:.\label{eq:click projection}
\end{equation}
The expectation of this for $c_{j}=1$ is the first-order click correlation
moment, $\left\langle \hat{\pi}_{j}\left(1\right)\right\rangle $,
which is the probability of observing a click at the $j$-th detector.
The projection operator for a specific binary pattern $\boldsymbol{c}$
is $\hat{\Pi}\left(\bm{c}\right)=\bigotimes_{j=1}^{M}\hat{\pi}_{j}\left(c_{j}\right)$,
whose expectation value is the Torontonian function \citep{quesada2018gaussian}.
This is exponentially small in almost all cases, which means it cannot
be measured for large scale experiments due to experimental sampling
errors.

To compute output probabilities of bosonic networks with threshold
detectors \textit{without} directly generating discrete patterns we
use grouped count probabilities (GCPs). These generate moments of
multiple possible output patterns. They also allow one to carry out
exponentially many high-order correlation tests. 

A GCP computes the probability of observing $\boldsymbol{m}=\left(m_{1},\dots,m_{d}\right)$
grouped counts in $d$-dimensions. Each grouped count $m_{j}$ is
obtained by summing over all binary patterns. These are combined into
bins based on the number of detector counts in a subset $S_{j}$ of
all $M$ output modes, such that $m_{j}=\sum_{i}^{M}c_{i}$. A $d$-dimensional
GCP is therefore defined as \citep{drummondSimulatingComplexNetworks2022}

\begin{equation}
\mathcal{G}_{\boldsymbol{S}}^{(n)}\left(\boldsymbol{m}\right)=\left\langle \prod_{j=1}^{d}\left[\sum_{\sum c_{i}=m_{j}}\hat{\Pi}_{S_{j}}\left(\boldsymbol{c}\right)\right]\right\rangle ,\label{eq:grouped prob}
\end{equation}
where $n=\sum_{j=1}^{d}M_{j}\leq M$ is the total click correlation
order, following Glauber's definition \citep{Glauber1963_CoherentStates},
and $\boldsymbol{S}=\left(S_{1},S_{2},\dots\right)$ is the vector
of disjoint subsets of $\boldsymbol{M}=\left(M_{1},M_{2},\dots\right)$
output modes. These includes marginal probabilities where some detectors
are not monitored, as well as moments like the Torontonian. However,
a GCP has the advantage of being both measurable and including data
from all detectors if required.

\subsubsection{Numerical computation}

As well as being unmeasurable, the Torontonian is not computable at
large scale. There are no efficient direct techniques, and phase-space
methods are only useful where the Torontonian has a large enough value
to exceed the theoretical sampling error. However, many GCPs are both
measurable and computable, as they are easily computed using the positive-P
representation. 

Due to the operator equivalence Eq.(\ref{eq:+P op equiv}), the the
normally ordered projection operator $\hat{\pi}_{j}$ is computed
via a replacement with the positive-P observable

\begin{equation}
\pi_{j}=e^{-n'_{j}}\left(e^{n'_{j}}-1\right)^{c_{j}},\label{eq:phase-space_click}
\end{equation}
where $n'_{j}$ is sampled from the output distribution Eq.(\ref{eq:output P}). 

The summation over detector outputs is efficiently carried out using
the multi-dimensional inverse discrete Fourier transform \citep{drummondSimulatingComplexNetworks2022}
\begin{equation}
\mathcal{G}_{\boldsymbol{S}}^{(n)}\left(\boldsymbol{m}\right)=\frac{1}{\prod_{j}(M_{j}+1)}\sum_{\boldsymbol{k}}\tilde{\mathcal{G}}_{M}^{(n)}(\boldsymbol{k})e^{i\sum_{j}k_{j}\theta_{j}m_{j}},
\end{equation}
where 

\begin{equation}
\mathcal{\tilde{G}}_{\boldsymbol{S}}^{(n)}\left(\boldsymbol{k}\right)=\left\langle \prod_{j=1}^{d}\bigotimes_{i\in S_{j}}\left(\pi_{i}(0)+\pi_{i}(1)e^{-ik_{j}\theta_{j}}\right)\right\rangle _{P},
\end{equation}
is the Fourier observable, $\theta_{j}=2\pi/(M_{j}+1)$ is the Fourier
angle, $k_{j}=0,\dots,M_{j}$ and $j=1,\dots,d$ defines the dimension. 

The Fourier transform is not only numerically efficient and highly
scalable, but allows all possible correlations present in the network
to be simulated, removing all patterns that don't contain $\boldsymbol{m}$
counts.

\subsubsection{Multi-dimensional binning of grouped correlations}

The experimentally reported total count probability \citep{zhong2020quantum,zhongPhaseProgrammableGaussianBoson2021},
which is the probability of observing $m$ clicks in any pattern,
is typically one of the first comparison tests experimental samples
are subjected to. It allows one to quickly determine whether outputs
are close to the expected distribution obtained using pure squeezed
state inputs, typically called the 'ideal' or 'ground truth' distribution
in the literature. 

However, Villalonga et al \citep{villalonga2021efficient} has shown
that total count distributions can be spoofed by classical algorithms
which sample from low-order marginal probabilities. Their second and
third-order samplers generate distributions that are closer to the
theoretical ideal distribution, approximated as a Gaussian in the
limit of large numbers of clicks, than an experiment which has claimed
quantum advantage \citep{zhongPhaseProgrammableGaussianBoson2021}.
Therefore, comparison tests are needed which utilize the true high-order
correlations generated in a linear network, to help differentiate
output distributions from experiments and classical sampling algorithms. 

This is where grouped probabilities with dimension $d>1$ become particularly
useful for statistical comparisons. A $d>1$-dimensional grouped correlation
of order $n=M$ is the probability of observing $m_{1},\dots,m_{d}$
grouped counts in the subsets 
\begin{equation}
\boldsymbol{S}=\left(S_{1},\dots,S_{d}\right)=\left(\{1,\dots,M/d\},\dots,\{M/d+1,\dots M\}\right),
\end{equation}
 such that $m_{1}=\sum_{i=1}^{M/d}c_{i}$, $m_{d}=\sum_{i=M/d+1}^{M}c_{i}$. 

The increased dimension means high-order correlations present in the
data become more statistically significant. This has two fundamental
benefits over one-dimensional comparisons; First, the increased dimension
generates a larger number of bins, or data points, that are available
for statistical testing. The number of bins generated per dimension
scales as $(M/d+1)^{d}$, although only a subset of these are used
for statistical testing (see Appendices). 

This allows a fine tuned comparison of theoretical and experimental
outputs. If statistical tests show discrepancies increase as dimension
increases, even after simple decoherence effects are included, this
could indicate further imperfections are affecting the network. 

Second, multi-dimensional GCPs simulated in phase-space provide an
efficient method for differentiating between data spoofed by low-order
sampling algorithms \citep{villalonga2021efficient,ohSpoofingCrossEntropy2022a}
and data generated from quantum experiments. This is due to such spoofing
algorithms generating patterns with an inherent bias as spoofing high-order
correlations generated in large size experiments is computationally
challenging for low-order sampling algorithms. 

Additional tests can be performed by randomly permuting each binary
pattern. This changes the output modes contained within each subset
$S_{j}$, leading to different values of $m_{j}$ for each permutation.
Without repetitions, there are 

\begin{equation}
\frac{\binom{M}{M/d}}{d}=\frac{M!}{d(M/d)!(M-M/d)!},
\end{equation}
possible ways of computing $m_{1},\dots,m_{d}$. 

This produces exponentially many non-trivial, randomized high-order
tests when $d>1$, allowing exponentially many comparisons to take
place, with different high-order correlations being observed in each
test.

If repeated comparisons show differences between theoretical and experimental
outputs remain statistically significant, one can hypothesize experimental
imperfections are causing samples to become inaccurate. These random
permutation tests can be simulated efficiently by applying the same
permutation used on the experimental samples to the rows of the transmission
matrix $\boldsymbol{T}$ used in the phase-space simulation. 

Theoretically, one can bin counts up to the maximum dimension possible
of $d=M$, which corresponds to the Torontonian. However, this is
strongly restricted by experimental sampling errors which increase
with dimension $d$ due to each bin containing progressively fewer
photon counts.

\subsubsection{Low-order click correlations}

Low-order marginal probabilities compute correlations over $n$ output
modes whilst ignoring the other $M-n$ modes \citep{renemaMarginalProbabilitiesBoson2020,renemaSimulabilityPartiallyDistinguishable2020},
with the number of observable correlations scaling as $\binom{M}{n}$.
For click detectors, the first and second-order click correlations
are defined as $\left\langle \hat{\pi}_{j}(1)\right\rangle $ and
$\left\langle \hat{\pi}_{j}(1)\hat{\pi}_{k}(1)\right\rangle $, respectively. 

The computational efficiency of computing low-order marginals of the
ideal output distribution has two advantages for GBS validation; It
allows a fast and direct comparison of specific experimental correlations
generated from a network with their expected values \citep{zhongPhaseProgrammableGaussianBoson2021},
and has formed foundation of the spoofing algorithms implemented in
Ref.\citep{villalonga2021efficient}, which compute the correct connected
correlations, also known as cumulants, of the ideal distribution for
orders $n\leq3$. 

The first two cumulants are defined as \citep{villalonga2021efficient,zhongPhaseProgrammableGaussianBoson2021,gardiner2004quantum}:

\begin{align}
\kappa_{1} & =\left\langle \hat{\pi}_{j}(1)\right\rangle \nonumber \\
\kappa_{2} & =\left\langle \hat{\pi}_{j}(1)\hat{\pi}_{k}(1)\right\rangle -\left\langle \hat{\pi}_{j}(1)\right\rangle \left\langle \hat{\pi}_{k}(1)\right\rangle ,
\end{align}
and describe the mean click count rate and covariance, respectively.
These low-order correlations are used by spoofing algorithms to generate
discrete binary patterns for large mode numbers without actually sampling
from the full Torontonian distribution. 

Computing moments of click correlations is efficient using GCPs simulated
in phase-space and is easily illustrated. For example, the third-order
click correlation is obtained by setting $n=3$, $S=\{j,k,h\}$ and
$\boldsymbol{m}=3$ where $\mathcal{G}_{\{j,k,h\}}^{(3)}(3)=\left\langle \hat{\pi}_{j}(1)\hat{\pi}_{k}(1)\hat{\pi}_{h}(1)\right\rangle $
is the probability of observing clicks at detectors $j,k,h$. 

\section{Scaling properties\label{sec:Sampling-errors-=000026}}

Linear networks produce sampled outputs with random observed photon
counts. Statistical testing is vital to determine the accuracy of
experimental samples \citep{Rukhin2010}. These statistical comparisons
also require an analysis of theoretical sampling errors, which should
be comparable or smaller than experimental sampling errors. Hence,
the scaling of computational cost or time is determined by the required
sample numbers.

Useful comparison simulations for validation and testing purposes
must be accurate. In this section, we use Glauber intensity correlations
to demonstrate how sampling errors grow with correlation order, illustrating
the importance of choosing the correct phase-space representation
to simulate bosonic networks with normally ordered detectors. We also
give an overview of the statistical estimates used throughout this
paper for the experimental and theoretical data.

Details of chi-squared and Z-statistic tests used to compare theory
with experiment are given in the Appendices.

\subsection{Experimental sampling errors}

We first analyze experimental sampling errors, which are crucial to
comparisons of theory and experiment. As the scientific issue is whether
theory and experiment agree, we must know what errors exist in the
data. We can only conclude there is a difference between theory and
experiment if a discrepancy cannot be explained by experimental or
theoretical sampling errors.

For any measurement, one would prefer theoretical errors less than
the experimental errors. However, there is little to be gained from
improving theoretical errors far below experimental errors, which
will give negligible additional information about the agreement of
theory and experiment. This is also relevant to computational complexity,
as explained below.

For the case of measured probabilities $P_{i}^{e}$ of an experimental
observation labeled $i$, given $x_{i}$ observations of the event
$i$ out of $N_{E}$ experimental samples \citep{pearson1900x}, the
estimated probability is obtained as
\begin{equation}
P_{i}^{e}=x_{i}/N_{E}.\label{eq:experiment_probability}
\end{equation}
For $x_{i}\gg1$, the variance in $P_{i}^{e}$ is $\sigma_{i}^{2}=P_{i}/N_{E}$,
where $P_{i}$ is the true underlying probability. 

To understand the scaling issues quantitatively, consider a typical
GBS experiment which generates $N_{E}$ random binary numbers $\bm{c}=\left[c_{1},\ldots,c_{M}\right]$.
Assuming $\left\langle \hat{\pi}_{j}(1)\right\rangle \approx\left\langle \hat{\pi}_{j}(0)\right\rangle $,
there are $2^{M}$ possible outcomes, and in recent large-scale experiments,
$2^{M}\gg N_{E}$. Data analyzed here has $N_{E}=4\times10^{7}$ experimental
binary patterns, with the full number of possible outcomes being $2^{144}=10^{43.3}$.

\textit{Hence, almost all patterns are never observed, even with $10^{7}$
samples.} A rough estimate is that $P_{i}\approx10^{-43.3}$ for a
``typical'' pattern. This is the average probability given by the
Torontonian function, which is known to be an exponentially hard ($\#P$)
function to compute. 

However, even in the special cases where it is computable, its not
useful for validation. The reason for this is that $x_{i}\gg1$ is
not satisfied, since in almost all cases one has $x_{i}=0$. Even
if the theoretical Torontonian could be computed to more than $43$
decimals, and the results stored, the experimental data would have
too large an error for useful comparisons. This is the reason why
verification requires binning to obtain significant experimental data.

Given a binning method such that there are less than $N_{E}$ distinct
outcomes $i$, one can obtain experimental probabilities with a relative
error that scales as $1/\sqrt{N_{E}}$. The goal of a theoretical
validation is to have an estimate for $P_{i}$ that can reach this
level of relative error in the binned Torontonian, even though the
Torontonian itself is exponentially hard to compute.

\subsection{Phase-space sampling error }

The computational process for simulating phase-space representations
is similar for any representation. First, samples $\alpha,\beta$
are generated by randomly sampling the input distribution $E_{S}$
times. For linear networks, the number of initial random numbers required
is proportional to $NE_{S}$ with a normally ordered method, or to
$ME_{S}$ with non-normally ordered methods, due to the additional
algebraic terms which arise from vacuum noise. 

If one is interested in dynamical simulations, the samples are then
propagated through time to solve a stochastic differential equation
\citep{carter1987squeezing}, the form of which changes depending
on the representation, the system Hamiltonian of interest and whether
losses are taken into account. However, we are only interested in
sampling from the output distribution, which is obtained by transforming
the input states as described above. 

Regardless of how the initial samples are transformed, output observables
are obtained in the form of a stochastic average over the entire ensemble
of samples. Therefore, the computation of the product of $E_{S}$
randomly sampled normally ordered output photon numbers $((n'_{j})^{c_{j}})^{(k)}$
is 

\begin{equation}
\bar{G}^{(n)}=\frac{1}{E_{S}}\sum_{k}^{E_{S}}((n'_{j})^{c_{j}})^{(k)}\dots\:((n'_{M})^{c_{M}})^{(k)},\label{eq:stochastic average}
\end{equation}
where the superscript $k$ denotes the label of a stochastic trajectory
in the overall ensemble, and $\bar{G}^{(n)}$ denotes the ensemble
mean. 

This is valid with all orderings if re-ordered to normal order, provided
the appropriate corrections are applied, and there are no correlation
terms with $c_{j}>1$. For normal ordering the terms can be repeated,
and the result is not restricted to the unitary case, since losses
can be included. In other cases, losses require additional noise terms.
The other orderings also introduce additional algebraic terms if there
are terms with $c_{j}>1$.

Stochastic averages are estimates of the actual theoretical probability
obtained from a quantum expectation value of an observed operator.
In the limit $E_{S}\rightarrow\infty$, ensemble means converge to
the actual theoretical probability such that in the case of Eq.(\ref{eq:stochastic average}),
$G^{(n)}=\lim_{E_{S}\rightarrow\infty}\bar{G}^{(n)}$. 

Practical implementations of phase-space ensemble averages typically
split ensembles into two sub-ensembles, so that $E_{S}=N_{S}N_{R}$
\citep{opanchuk2018simulating}. This has a computational advantage,
allowing efficient vector and multi-core parallel computing, and reducing
time requirements for large ensemble sizes. There is also a statistical
benefit: the first sub-ensemble $N_{S}$ is the number of samples
of the initial state. For $N_{S}\rightarrow\infty$, this gives sample
averages that are normally distributed via the central limit theorem. 

The second sub-ensemble $N_{R}$ is the number of times the computation
is repeated. This is equivalent to sampling from a normal distribution
$N_{R}$ times \citep{opanchuk2018simulating}. Therefore, the actual
computation of the stochastic average Eq.(\ref{eq:stochastic average})
proceeds as 

\begin{equation}
\bar{G}^{(n)}=\frac{1}{N_{R}}\sum_{i=1}^{N_{R}}\left(\frac{1}{N_{S}}\sum_{h=1}^{N_{S}}((n'_{j})^{c_{j}})^{(h)}\dots\:((n'_{M})^{c_{M}})^{(h)}\right)_{(i)},\label{eq:stochastic average sub-ens}
\end{equation}
where $h$, $i$ are the number of samples of the first and second
sub-ensembles, respectively. 

The second sub-ensemble also generates a statistical estimate of the
theoretical sampling error of the ensemble mean as $\sigma_{T}=\sigma_{t}/\sqrt{N_{R}}$,
where the sub-ensemble variance is \citep{kloedenStochasticDifferentialEquations1992,freundStatisticalMethods2003}:

\begin{equation}
\sigma_{t}^{2}=\frac{\sum_{i=1}^{N_{R}}\left(\bar{G}_{(i)}^{(n)}-\bar{G}^{(n)}\right)^{2}}{N_{R}-1},
\end{equation}
where we define the sub-ensemble mean as the $i$-th sum over the
simulated data:
\begin{equation}
\bar{G}_{(i)}^{(n)}=\left(\frac{1}{N_{S}}\sum_{h=1}^{N_{S}}((n'_{j})^{c_{j}})^{(h)}\dots\:((n'_{M})^{c_{M}})^{(h)}\right).
\end{equation}
Thus, the theoretical standard deviation in the mean for $\bar{G}^{(n)}$
is readily obtained from the simulated fluctuations in $\bar{G}_{(i)}^{(n)}$.

A computationally friendly definition of $\sigma_{t}$ can be derived
which doesn't require computing $\bar{G}^{(n)}$ before performing
the summation using the expansion $\sum_{i=1}^{N_{R}}\left(\bar{G}_{(i)}^{(n)}-\bar{G}^{(n)}\right)^{2}=\sum_{i=1}^{N_{R}}\left(\bar{G}_{(i)}^{(n)}\right)^{2}-\left(\sum_{i=1}^{N_{R}}\bar{G}_{(i)}^{(n)}\right)^{2}/N_{R}$
\citep{freundStatisticalMethods2003}. 

Therefore, when $N_{R}\gg1$, the theoretical sampling error of the
correlation of $E_{S}$ randomly sampled output photon numbers is
estimated using the computationally efficient form

\begin{equation}
\sigma_{T}=\sqrt{\frac{\sum_{i=1}^{N_{R}}\left(\left(\bar{G}_{(i)}^{(n)}\right)\right)^{2}-\left(\sum_{i=1}^{N_{R}}\left(\bar{G}_{(i)}^{(n)}\right)\right)^{2}/N_{R}}{N_{R}\left(N_{R}-1\right)}}.\label{eq:theo samp error}
\end{equation}
As with any sampling procedure, the sampling error can be reduced
by increasing the total number of ensembles which corresponds to increasing
either sub-ensemble. Increasing $N_{S}$ requires more memory and
processing power, while the speed of an increased $N_{R}$ depends
on whether multi-core computing is possible. The more cores available
the faster the computation runs.

\subsection{Scaling properties of phase-space sampling errors}

To quantify the scaling properties of different phase-space methods,
intensity correlations with increasing order are computed using the
Wigner, Q and positive-P representations. We consider an $M=20$ mode
bosonic network with unit transmission matrix and $N=20$ input pure
squeezed states with a uniform squeezing parameter of $\boldsymbol{r}=[1,\dots,1]$.
For a network of this type, the output intensity correlations can
be computed exactly. Therefore, one can use the ratio of theoretical
sampling errors, estimated by Eq.(\ref{eq:theo samp error}), and
exactly computed correlations with increasing order to determine how
sampling errors of each representation grow with correlation order. 

Comparisons are plotted in Fig.(\ref{fig:sampling error growth})
for simulations with $E_{S}=4.8\times10^{7}$ ensembles. The Q and
Wigner representations produce sampling errors many orders of magnitude
larger than positive-P simulations, and become approximately equal
to the computed intensity correlations at orders $n=6$ and $n=8$,
respectively. At this point, the numerical sampling error is too large
for useful results.

By contrast, the positive-P sampling error is always much smaller,
with growth as $n$ increases arising from sampling distributions
with decreasing probabilities. These are also increasingly difficult
to measure. 

This shows the benefit of the normally-ordered approach, which gives
exponentially lower error with increasing order compared to other
types of ordering. Because we are mostly interested in high-order
correlations, we do not use the non-normally ordered methods further. 

In earlier work we compared theoretical phase-space simulation errors
with expected experimental sampling errors for number-state boson
sampling, and found that the theoretical sampling variances are typically
much lower than the experimental variances for the same number of
samples \citep{opanchuk2018simulating}. This scaling persists in
the case of GBS, with detailed results to be given elsewhere.

\subsection{Summary of sampling errors}

In summary, it is exponentially hard to estimate probabilities in
the unbinned, high order limit, but it is also exponentially hard
to measure the probabilities, because in both cases exponentially
many samples would be needed. Once the data is binned or marginalized
so that it becomes measurable, we find that the positive-P method
gives exponentially lower sampling error than other phase-space methods
that do not use normal-ordering.

In a previous paper \citep{opanchuk2018simulating}, it was shown
that the phase-space method also has exponentially lower errors than
experiment for identical sample numbers. Hence we are generally able
to employ lower numbers of samples than were used in the experiment,
which gives a reasonable benchmark for timing comparisons. Apart from
this, the timing of the simulations is polynomial, not exponential
in the mode number.

\section{Comparison of theory and experiment\label{sec:Comp_theory_exp}}

In this section, we compare theoretical GCPs with experimental data
from a $144$-mode GBS linear network \citep{zhongPhaseProgrammableGaussianBoson2021}.
This experiment obtained data for two different laser waists, $125\mu m$
and $65\mu m$, and varying laser power. The first waist contains
data for two different powers and the second waist, five different
powers. Squeezing parameters are $50$-mode vectors of amplitude $\boldsymbol{r}$,
one for each laser power tested, while the transmission matrices $\boldsymbol{T}$
are of size $50\times144$ with two matrices in total, one for each
laser waist. 

Test statistics for comparisons of GCPs and first-order click correlation
moments for all data sets are given, however comparison plots are
also only shown for data from laser waist $65\mu m$ and power $1.65W$.
This is due to both practicality and claims of computational advantage
for this data set, as the cost of computing the Torontonian and generating
random outputs scales with the number of modes and hence detector
clicks. 

This experimental data was compared with an ideal squeezed state model
based on the reported experimental parameters, and the thermalized
squeezed state model described in Subsection (\ref{subsec:Input-state}),
with values of $\epsilon,$$t$ chosen to minimize the errors in the
total count distributions. 

This model does not account for possible inhomogeneities in thermalization
for different inputs, nor for inhomogeneities in output transmission
or detector efficiency.

Raw data from the tested experiment can be found from \citep{RawDataJiuzhang}.
Extracted data used for comparisons in this work, as well as software
for generating the phase-space simulations, can be found in the fully
open source and documented software package developed for simulating
linear bosonic networks with threshold detectors in phase-space, called
xQSim, which is available from \citep{GitHubPeterddrummondXqsim}.

\subsection{Total count distributions}

\begin{table*}
\begin{centering}
\begin{tabular}{cccccccccc}
\toprule 
\multirow{2}{*}{Waist} & \multirow{2}{*}{Power} & \multicolumn{3}{c}{Pure squeezed inputs} & \multirow{2}{*}{$t$} & \multirow{2}{*}{$\epsilon$} & \multicolumn{3}{c}{Thermalized squeezed inputs}\tabularnewline
\cmidrule{3-5} \cmidrule{4-5} \cmidrule{5-5} \cmidrule{8-10} \cmidrule{9-10} \cmidrule{10-10} 
 &  & $\chi_{EI}^{2}/k$ & $k$ & $Z_{EI}$ &  &  & $\chi_{ET}^{2}/k$ & $k$ & $Z_{ET}$\tabularnewline
\midrule
\midrule 
\multirow{2}{*}{$125\mu m$} & $1.412W$ & $218$ & $53$ & $78$ & $0.9972$ & $0.0354$ & $2\pm0.5$ & $52$ & $4\pm1.5$\tabularnewline
\cmidrule{2-10} \cmidrule{3-10} \cmidrule{4-10} \cmidrule{5-10} \cmidrule{6-10} \cmidrule{7-10} \cmidrule{8-10} \cmidrule{9-10} \cmidrule{10-10} 
 & $0.5W$ & $143$ & $31$ & $50$ & $1.0006$ & $0.0392$ & $20\pm2$ & $31$ & $20\pm1$\tabularnewline
\midrule 
\multirow{5}{*}{$65\mu m$} & $1.65W$ & $1861$ & $85$ & $221$ & $1.0109$ & $0.0428$ & $10\pm1$ & $84$ & $23\pm2$\tabularnewline
\cmidrule{2-10} \cmidrule{3-10} \cmidrule{4-10} \cmidrule{5-10} \cmidrule{6-10} \cmidrule{7-10} \cmidrule{8-10} \cmidrule{9-10} \cmidrule{10-10} 
 & $1W$ & $215$ & $74$ & $91$ & $1.0026$ & $0.0354$ & $6\pm1$ & $73$ & $15\pm2$\tabularnewline
\cmidrule{2-10} \cmidrule{3-10} \cmidrule{4-10} \cmidrule{5-10} \cmidrule{6-10} \cmidrule{7-10} \cmidrule{8-10} \cmidrule{9-10} \cmidrule{10-10} 
 & $0.6W$ & $171$ & $57$ & $72$ & $0.9966$ & $0.0288$ & $2.5\pm1$ & $57$ & $6\pm2$\tabularnewline
\cmidrule{2-10} \cmidrule{3-10} \cmidrule{4-10} \cmidrule{5-10} \cmidrule{6-10} \cmidrule{7-10} \cmidrule{8-10} \cmidrule{9-10} \cmidrule{10-10} 
 & $0.3W$ & $193$ & $40$ & $64$ & $0.9972$ & $0.0202$ & $7\pm1$ & $40$ & $12\pm1$\tabularnewline
\cmidrule{2-10} \cmidrule{3-10} \cmidrule{4-10} \cmidrule{5-10} \cmidrule{6-10} \cmidrule{7-10} \cmidrule{8-10} \cmidrule{9-10} \cmidrule{10-10} 
 & $0.15W$ & $151$ & $28$ & $49$ & $0.9972$ & $0.0208$ & $1.2\pm0.3$ & $27$ & $0.7\pm1$\tabularnewline
\bottomrule
\end{tabular}
\par\end{centering}
\caption{Statistical test outputs for comparisons of total counts $\mathcal{G}_{144}^{(144)}(m)$
for all data sets obtained from a $144$-mode GBS experiment \citep{zhongPhaseProgrammableGaussianBoson2021}.
Chi-square and approximate Z-statistic tests, defined analytically
in the appendix, are generated from comparisons with phase-space simulations
of $E_{S}=1.2\times10^{6}$ ensembles. Comparisons of experiment with
simulations of pure squeezed state inputs, the ideal GBS, are denoted
by the subscript $EI$, whilst comparisons of simulated thermalized
squeezed states are denoted by the subscript $ET$. Parameters $t$
and $\epsilon$ represent corrections to the experimental $\boldsymbol{T}$
matrix and the thermal component added to the input states, respectively.
The chi-square and Z-statistic error bars are due to uncertainties
in the corresponding $t,\epsilon$ values. The fitting parameters
for each data set had error bars for $t,\epsilon$ of $\pm0.0005$.
\label{tab:Total counts chi-square}}
\end{table*}

We first present comparisons of the experimentally reported total
counts, which is computed as a $d=1$ dimensional GCP. Simulations
of pure squeezed state inputs, corresponding to the ideal GBS, are
compared to binned experimental data in Table. \ref{tab:Total counts chi-square}
for each available data set, with the distance between distributions
signified by the subscript $EI$. Probability distance measures are
defined analytically in the Appendix. 

Since linear networks do not change the Gaussian nature of the input
state, the output state will be Gaussian and one expects $\chi_{EI}^{2}/k\approx1$.
This is clearly not the case for the reported experimental data, with
$\chi_{EI}^{2}/k\gg1$ for all experimental samples. 

Although data set $65\mu m$, $1.65W$ has the largest $\chi_{EI}^{2}/k$
output of $\chi_{EI}^{2}/k\approx1.9\times10^{3}$ for $k=85$, Z-statistical
tests indicate data from all experiments are far from their expected
normally distributed mean for pure squeezed state inputs. 

\begin{figure*}
\begin{centering}
\includegraphics[width=0.9\textwidth]{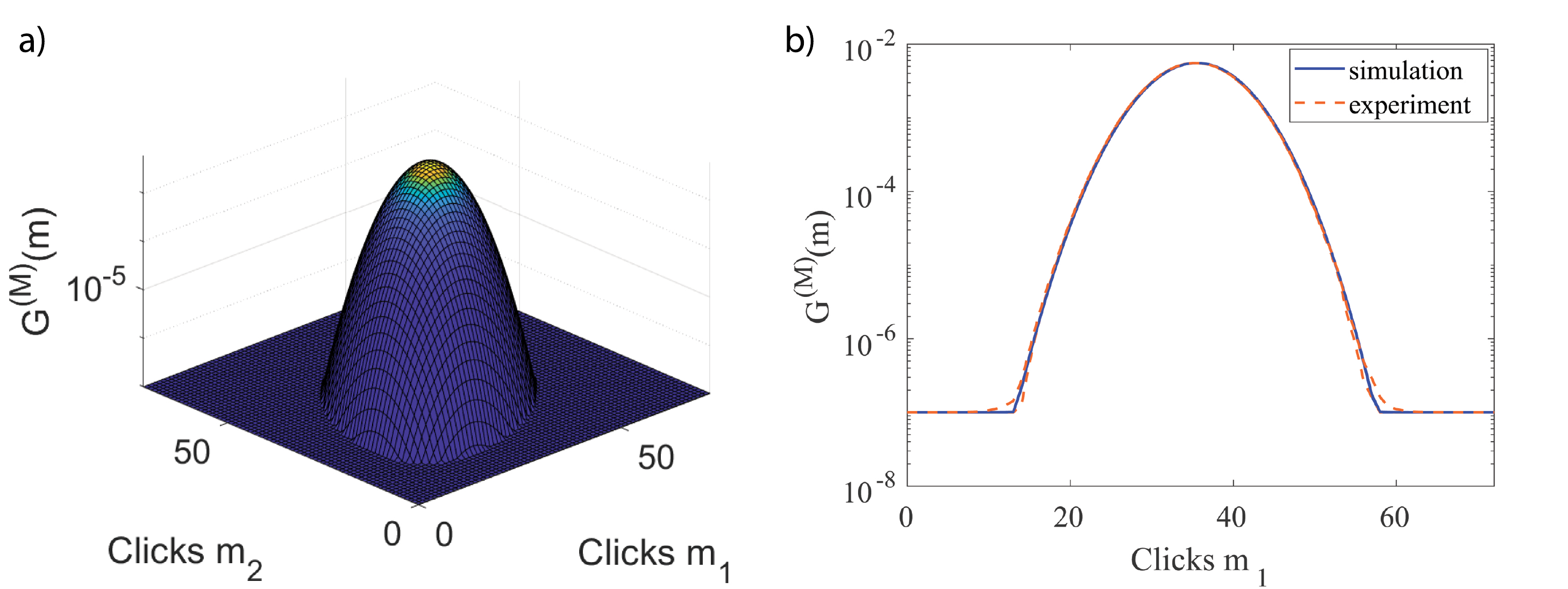}
\par\end{centering}
\caption{Comparisons of experimental data set $65\mu m$, $1.65W$ with simulations
of $1.2\times10^{6}$ stochastic ensembles for a $d=2$ dimensional
GCP, $\mathcal{G}_{72,72}^{(144)}(m_{1},m_{2})$, with input decoherence.
a) Full two-dimensional comparison distribution of GCPs with all $73^{2}$
data points. b) One-dimensional slice through the maximum of the two-dimensional
distribution. Plots a comparison of grouped count $m_{1}$ where the
solid blue line is the theoretical prediction and the orange dashed
line is the experimental data. \label{fig:2D binning}}
\end{figure*}

\begin{figure}
\begin{centering}
\includegraphics[width=0.95\columnwidth]{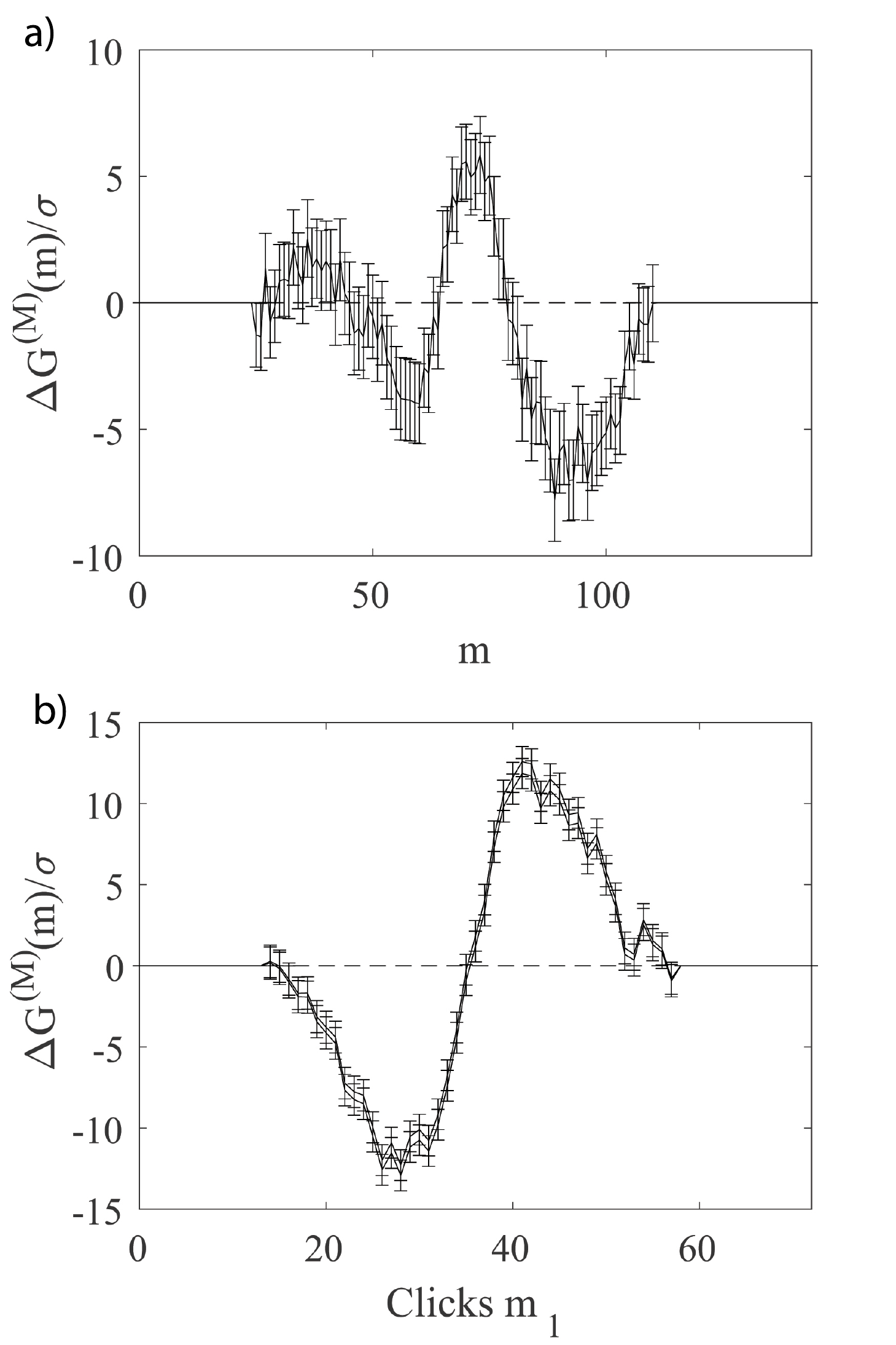}
\par\end{centering}
\caption{Normalized difference of experiment versus simulation for data set
$65\mu m$, $1.65W$ with added decoherence. a) Plots results of $\Delta\mathcal{G}^{(M)}(m)/\sigma_{m}$
versus $m$ for total count probability distributions. b) Results
are $\Delta\mathcal{G}^{(M)}(m)/\sigma_{m}$ versus $m_{1}$ of the
one-dimensional slice shown in Fig.(\ref{fig:2D binning}).b which
is a $d=2$ dimensional GCP comparison. Upper and lower lines correspond
to theoretical sampling errors. \label{fig:Non-perm diff comp}}
\end{figure}

Improved agreement is obtained when decoherence is added to simulations
as shown in Table. \ref{tab:Total counts chi-square}, which contains
chi-square and Z-statistic outputs for each data set, with $ET$ denoting
the distance between experiment and thermalized distributions. Corresponding
fitting parameters are also given. 

After input decoherence corresponding to $\approx2\%$ mode mismatches
is included, the samples obtained from an experiment using laser waist
and power $65\mu m$, $0.15W$ are closer to the simulated distribution
than any other data set, as indicated by the result that $Z\approx0.7$.
Unlike the pure squeezed state comparisons, the Z-statistic shows
that detected total count distributions agree with the expected distribution
for input squeezed states with added decoherence. 

Such good agreement is not the case with most other data sets, where
Z values remain large. This is particularly the case with samples
corresponding to $65\mu m$, $1.65W$, which required the largest
amount of input decoherence, equating to $\approx4\%$ mode mismatch,
to obtain a three-orders of magnitude improvement in $\chi^{2}/k$
values compared to pure state inputs. Despite this improvement, the
Z-statistic shows the probability of obtaining such an output is very
small, indicating systematic errors.

\subsection{Two-dimensional binning comparisons}

\begin{table}
\begin{centering}
\begin{tabular}{cccccccc}
\toprule 
\multirow{2}{*}{Waist} & \multirow{2}{*}{Power} & \multicolumn{4}{c}{Two-dimensional GCP} & \multicolumn{2}{c}{Four-dimensional GCP}\tabularnewline
\cmidrule{3-8} \cmidrule{4-8} \cmidrule{5-8} \cmidrule{6-8} \cmidrule{7-8} \cmidrule{8-8} 
 &  & $Z_{EI}$ & $k$ & $Z_{ET}$ & $k$ & $Z_{ET}$ & $k$\tabularnewline
\midrule
\midrule 
\multirow{2}{*}{$125\mu m$} & $1.412W$ & $181$ & $712$ & $145$ & $702$ & $153$ & $28855$\tabularnewline
\cmidrule{2-8} \cmidrule{3-8} \cmidrule{4-8} \cmidrule{5-8} \cmidrule{6-8} \cmidrule{7-8} \cmidrule{8-8} 
 & $0.5W$ & $125$ & $287$ & $115$ & $285$ & $209$ & $6479$\tabularnewline
\midrule 
\multirow{5}{*}{$65\mu m$} & $1.65W$ & $422$ & $1582$ & $185$ & $1567$ & $200$ & $98682$\tabularnewline
\cmidrule{2-8} \cmidrule{3-8} \cmidrule{4-8} \cmidrule{5-8} \cmidrule{6-8} \cmidrule{7-8} \cmidrule{8-8} 
 & $1W$ & $168$ & $1274$ & $68$ & $1267$ & $109$ & $70826$\tabularnewline
\cmidrule{2-8} \cmidrule{3-8} \cmidrule{4-8} \cmidrule{5-8} \cmidrule{6-8} \cmidrule{7-8} \cmidrule{8-8} 
 & $0.6W$ & $107$ & $825$ & $32$ & $815$ & $66$ & $33271$\tabularnewline
\cmidrule{2-8} \cmidrule{3-8} \cmidrule{4-8} \cmidrule{5-8} \cmidrule{6-8} \cmidrule{7-8} \cmidrule{8-8} 
 & $0.3W$ & $105$ & $449$ & $43$ & $445$ & $99$ & $13605$\tabularnewline
\cmidrule{2-8} \cmidrule{3-8} \cmidrule{4-8} \cmidrule{5-8} \cmidrule{6-8} \cmidrule{7-8} \cmidrule{8-8} 
 & $0.15W$ & $76$ & $242$ & $40$ & $240$ & $111$ & $4695$\tabularnewline
\bottomrule
\end{tabular}
\par\end{centering}
\caption{Summary of Z-statistic test outputs for comparisons of GCPs with dimensions
$d=2$ and $d=4$ for all GBS experimental data sets obtained from
Ref.\citep{zhongPhaseProgrammableGaussianBoson2021}. Simulations
are performed with $E_{S}=1.2\times10^{6}$ ensembles for pure squeezed
states and thermalized squeezed states with the fitting parameters
used to obtain the minimized total counts $\chi^{2}/k$ outputs for
each data set obtained from Table. \ref{tab:Total counts chi-square}.
\label{tab:Summary_multi_dim_Z}}

\end{table}

To gain further insight into the experimental data, we present comparisons
of multi-dimensional GCPs with fitting parameters corresponding to
those given in Table. \ref{tab:Total counts chi-square} for experimental
samples of every tested laser waist and power. Due to the large number
of valid bins, the Z-statistic is the most useful statistical test
for multi-dimensional GCPs. The increased number of data points produces
Gaussian distributions with much smaller variances, meaning comparisons
are required to pass a more stringent test. Therefore, Z-statistic
outputs for multi-dimensional GCPs are presented in Table. \ref{tab:Summary_multi_dim_Z}. 

We start by analyzing comparisons of a $d=2$ dimensional GCP, an
example of which is plotted in Fig.(\ref{fig:2D binning}). Output
Z values increase for all data sets when compared to the corresponding
total count outputs of Table. \ref{tab:Total counts chi-square}.
This is particularly noticeable for data from $65\mu m$, $0.15W$,
which sees a forty-fold increase in the $Z$ value, with $Z_{ET}\approx40$,
when compared to total counts for simulations with added decoherence,
which have $Z_{ET}\approx0.7$.

Comparisons with simulations using pure squeezed state inputs record
even larger statistical errors, with data from $65\mu m$, $1.65W$
giving $Z_{EI}\approx422$ for $1582$ valid bins containing more
than $10$ counts. This is improved upon by adding decoherence in
the inputs, giving $Z_{ET}\approx185$, which is still very large. 

Therefore, statistical testing indicates that experimental samples
for all laser waists and powers are further away from the expected
multi-dimensional results than comparisons of total counts. 

This increased difference between theory and experiment for growing
dimension is reflected in Fig.(\ref{fig:Non-perm diff comp}), which
plots the normalized difference between theory and experiment \citep{drummondSimulatingComplexNetworks2022}: 

\begin{equation}
\frac{\Delta\mathcal{G}^{(M)}(m)}{\sigma_{m}}=\frac{\bar{\mathcal{G}}-\mathcal{G}^{e}}{\sigma_{m}}.
\end{equation}
We use the shorthand notation $\bar{\mathcal{G}}$ to denote the phase-space
simulated GCP ensemble mean and $\mathcal{G}^{e}$ to denote the experimental
GCP. The normalized difference follows from Eq.(\ref{eq:grouped count chi-square}),
except $\sigma_{m}$ only sums over theoretical and experimental variances
of the compared $m$ grouped counts. 

The finer comparison obtained from increased dimensional GCPs shows
experimental data has further, underlying differences that arise when
higher-order correlations are simulated. To confirm this is the case,
we can repeat the two-dimensional GCP comparison tests an exponential
number of times by randomly permuting each binary pattern. Although
not all of these tests can be performed, we randomly permute patterns
from $65\mu m$, $1.65W$ ten times and $125\mu m$, $1.412W$ five
times to determine whether differences remain significant. 

\begin{figure}
\begin{centering}
\includegraphics[width=0.95\columnwidth]{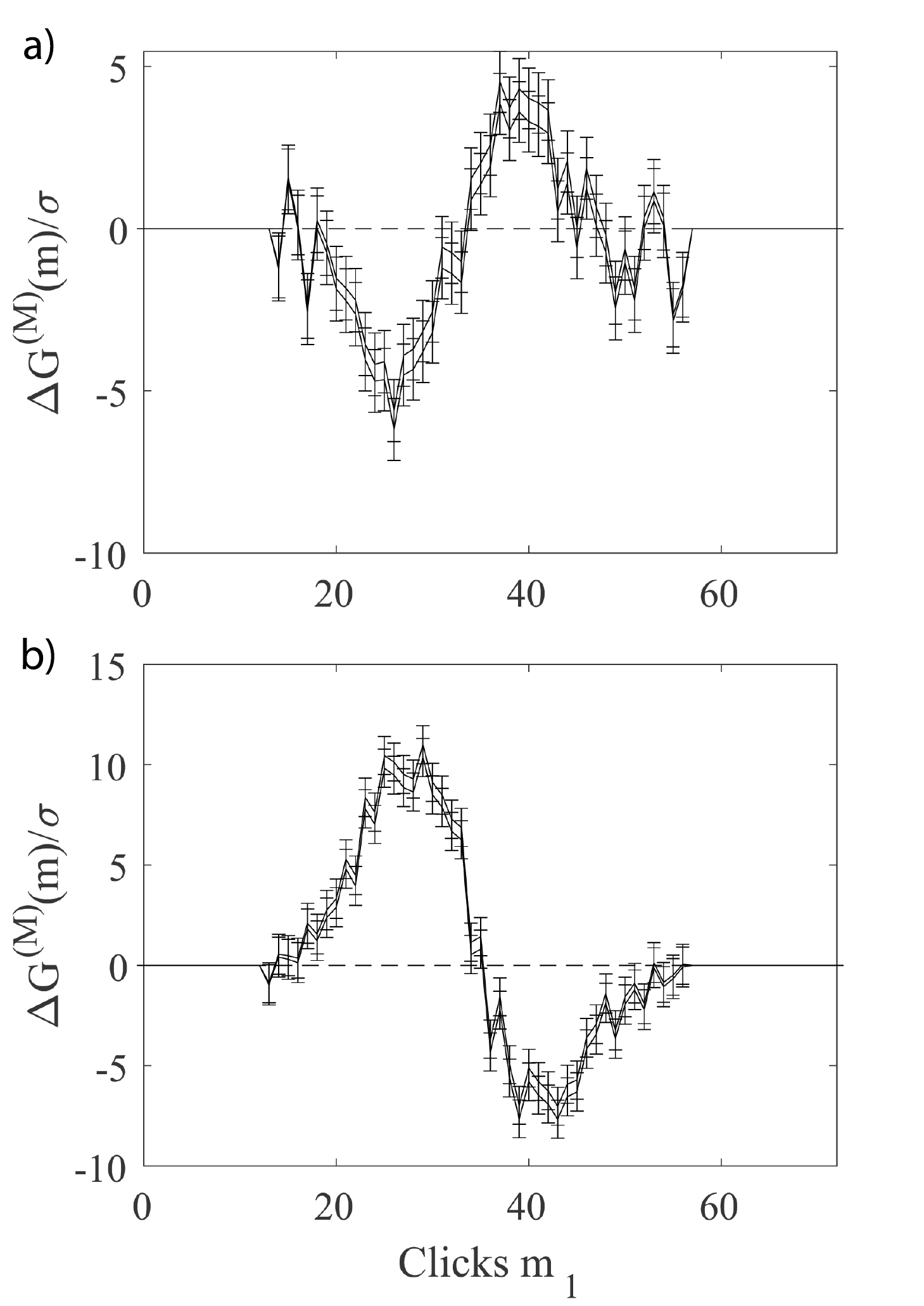}
\par\end{centering}
\caption{Comparison of normalized differences for two random permutations of
binary patterns from data set $65\mu m$, $1.65W$. The observable
is a $d=2$ dimensional GCP with input decoherence which is simulated
with $E_{S}=1.2\times10^{6}$. Graphed results are of $\Delta\mathcal{G}^{(M)}(m)/\sigma_{m}$
versus $m_{1}$. a) Plots a random permutation test with a statistic
test output of $Z_{ET}\approx53$ with $k=1547$ valid bins, whilst
b) plots a random permutation test with $Z_{ET}\approx142$ for $k=1567$.
Upper and lower lines are $\pm1\sigma_{T,i}$ theoretical sampling
errors. \label{fig:Random perm diff plots}}
\end{figure}

Since each random permutation obtains a different value for the grouped
count $m_{j}$, comparison differences are likely to vary, with some
tests showing better agreement than others. This is seen in Fig.(\ref{fig:Random perm diff plots})
which compares the normalized difference of two random permutations,
out of a tested ten, for samples from $65\mu m$, $1.65W$. Although
the average Z value for all ten random permutations is $\left\langle Z_{ET}\right\rangle _{rp}\approx115$
for $\left\langle k\right\rangle _{rp}=1568$, where $\left\langle \dots\right\rangle _{rp}$
denotes averages over random permutations, these two permutations
obtain $Z_{ET}\approx53$ with $k=1547$ and $Z_{ET}\approx143$ for
$1567$ valid bins. Meanwhile five permutations of patterns from $125\mu m$,
$1.412W$ output an average of $\left\langle Z_{ET}\right\rangle _{rp}\approx94$
for $\left\langle k\right\rangle _{rp}=704$. 

Each random permutation performed on both data sets generates a mean
decrease in $Z$ test outputs when compared to non-permutation comparisons
presented in Table. \ref{tab:Summary_multi_dim_Z}. However, every
permutation is producing statistical outputs with exponentially small
probabilities of being observed.

Therefore, sample detector counts show likely departures from randomness
for not only these data sets, but possibly all data sets from this
experiment, when compared to both ideal and mode mismatched theoretical
distributions.

\subsection{Four-dimensional binning comparisons}

\begin{figure*}
\begin{centering}
\includegraphics[width=0.9\textwidth]{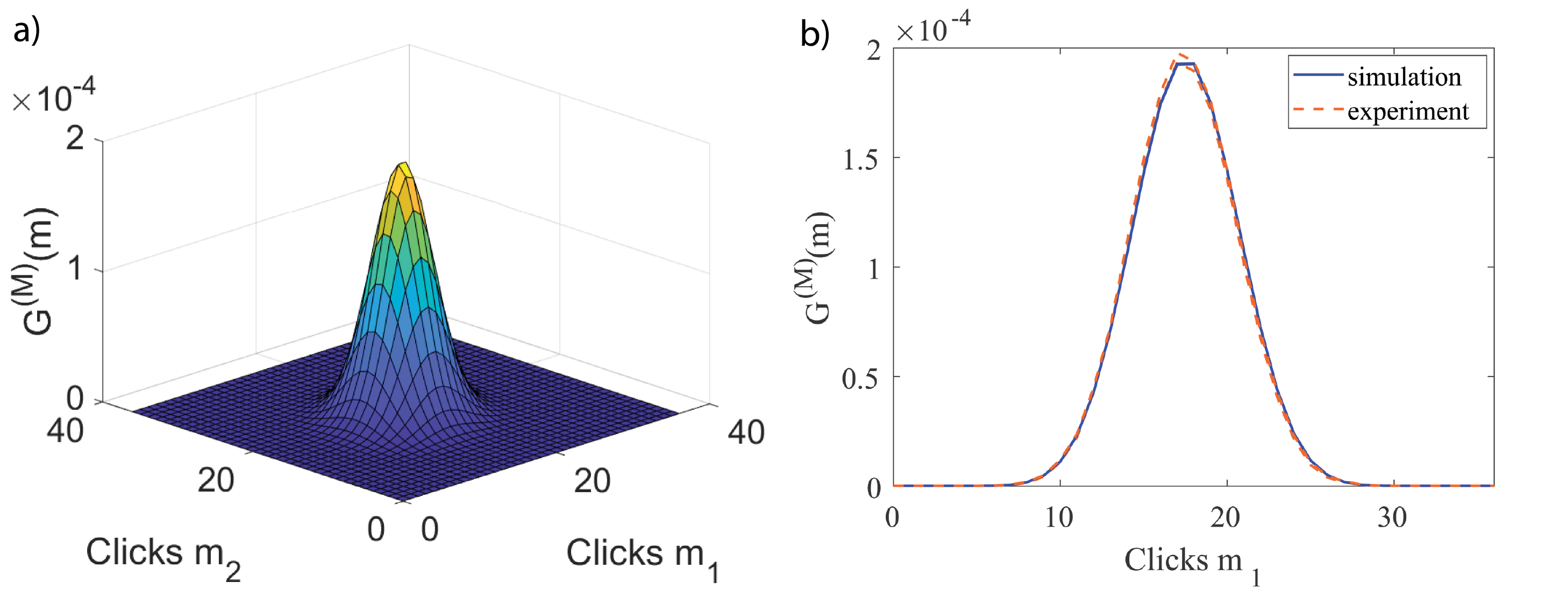}
\par\end{centering}
\caption{Comparisons of a $d=4$ dimensional GCP, $\mathcal{G}_{36,36,36,36}^{(144)}(m_{1},m_{2},m_{3},m_{4})$,
for data set $65\mu m$, $1.65W$ containing $N_{E}=4\times10^{7}$
experimental binary patterns and phase-space simulations with $E_{S}=1.2\times10^{6}$
and added decoherence. a) A two-dimensional planar slice of grouped
counts $m_{1},m_{2}$ of a four-dimensional distribution. All $37^{4}$
data points are used for statistical tests. b) One-dimensional slice
through the maximum of the two-dimensional planar slice. Plot is a
comparison of $m_{1}$, where the solid blue line is the theoretical
prediction and the orange dashed line is the experimental GCP.\label{fig:4D binning}}
\end{figure*}

To further test if the experimental data agrees with the Gaussian
model, we increased the dimension of the binning to $d=4$. Statistical
testing of outputs from each data set are given in Table. \ref{tab:Summary_multi_dim_Z}
and graphical comparisons of samples from $65\mu m$, $1.65W$ are
shown in Fig.(\ref{fig:4D binning}). This relatively moderate increase
in dimension causes a dramatic increase in the number of bins satisfying
$m_{j}>10$, giving more data points to test. 

At first glance, it appears as though the Z-statistic for some data
has stabilized, particularly $65\mu m$, $1.65W$ and $125\mu m$,
$1.412W$, where the large increases seen when going from one to two-dimensions
is not observed. This is further verified by performing comparisons
for five random permutations of both data sets, giving averages of
$\left\langle Z_{ET}\right\rangle _{rp}\approx159$ for $\left\langle k\right\rangle _{rp}=99650$
and $\left\langle Z_{ET}\right\rangle _{rp}\approx136$ for $\left\langle k\right\rangle _{rp}=29175$,
respectively. 

However, the chi-square and $Z$-statistic tests require summing over
both experimental and theoretical sampling errors (see Appendices).
Therefore, if either the mean theoretical, $\bar{\sigma}_{T}$, or
experimental, $\bar{\sigma}_{E}$, sampling errors satisfy $\bar{\sigma}_{E}\gg\bar{\sigma}_{T}$
or $\bar{\sigma}_{T}\gg\bar{\sigma}_{E}$ while $\bar{\sigma}_{E},\bar{\sigma}_{T}\approx\bar{\Delta}$,
where $\bar{\Delta}=\sum_{i}^{k}(\bar{\mathcal{G}}_{i}-\mathcal{G}_{i}^{e})/k$
is the mean difference error, an artificially small $\chi^{2}/k$
is generated causing the $Z$ value to appear to stabilize. 

A closer inspection shows experimental sampling errors for non-permutation
comparisons of $65\mu m$, $1.65W$ have a mean value of $\bar{\sigma}_{E}\approx3.6\times10^{-7}$.
This is not only much larger than theoretical sampling errors, $\bar{\sigma}_{T}\approx2.1\times10^{-8}$,
but also reaches the level where $\bar{\sigma}_{E}\approx\bar{\Delta}$.
This is also observed in sampling errors from $125\mu m$, $1.412W$
which satisfy $\bar{\sigma}_{E}\gg\bar{\sigma}_{T}$ for non-permutation
comparisons, with $\bar{\sigma}_{E}\approx5.4\times10^{-7}$ and $\bar{\sigma}_{T}\approx4.9\times10^{-8}$.

The cause of this increase in $\bar{\sigma}_{E}$ is due to experimental
GCPs containing many bins with few photon counts per bin. Therefore,
with currently available experimental data, experimental sampling
errors become significant at four-dimensions, rendering comparisons
less accurate. 

In other words, there is a balance required between test complexity
and sample numbers. While more complex tests are much harder to fake
because there are exponentially many of them, there is a price to
pay. The amount of experimental data required to give low experimental
sampling errors becomes unfeasibly large, reducing the power of the
tests. Despite this limitation, we note that for the four-dimensional
binning case, $Z_{ET}$ is still too large. 

This may indicate that the transmission errors change from mode to
mode, which we do not take into account here. 

\subsection{Low-order moments\label{subsec:Low-order-moments}}

\begin{table}
\begin{centering}
\begin{tabular}{cccc}
\toprule 
\multirow{1}{*}{Waist} & \multirow{1}{*}{Power} & $Z_{EI}$ & $Z_{ET}$\tabularnewline
\midrule
\midrule 
\multirow{2}{*}{$125\mu m$} & $1.412W$ & $285$ & $284$\tabularnewline
\cmidrule{2-4} \cmidrule{3-4} \cmidrule{4-4} 
 & $0.5W$ & $418$ & $425$\tabularnewline
\midrule 
\multirow{5}{*}{$65\mu m$} & $1.65W$ & $541$ & $516$\tabularnewline
\cmidrule{2-4} \cmidrule{3-4} \cmidrule{4-4} 
 & $1W$ & $405$ & $397$\tabularnewline
\cmidrule{2-4} \cmidrule{3-4} \cmidrule{4-4} 
 & $0.6W$ & $267$ & $263$\tabularnewline
\cmidrule{2-4} \cmidrule{3-4} \cmidrule{4-4} 
 & $0.3W$ & $235$ & $231$\tabularnewline
\cmidrule{2-4} \cmidrule{3-4} \cmidrule{4-4} 
 & $0.15W$ & $198$ & $196$\tabularnewline
\bottomrule
\end{tabular}
\par\end{centering}
\caption{Tabulated Z-statistic outputs for all $144$ possible first-order
click correlation moments of each data set. Simulations are performed
using $1.44\times10^{7}$ ensembles for input stochastic samples corresponding
to pure squeezed states with $\epsilon=0$ and $t=1$, corresponding
to the ideal GBS. The distance between experimental data and the simulated
ideal output denoted by the subscript $EI$, whilst $ET$ is the distance
between experiment and simulations with small admixtures of thermal
decoherence corresponding to mode mismatches. Fitting parameters applied
to each data set can be found in Table. \ref{tab:Total counts chi-square}.
\label{tab:Summary first-order click Z-stat}}
\end{table}

Previously, click correlation moments, in the form of cumulants, have
been used to verify the presence of non-trivial correlations in experimental
data and to determine the accuracy of spoofing algorithms \citep{zhongPhaseProgrammableGaussianBoson2021,villalonga2021efficient}.
Comparisons of these marginal distributions are useful in determining
whether specific experimental correlations agree with theoretical
probabilities. 

The simplest of these is the first-order moment, $\left\langle \hat{\pi}_{j}(1)\right\rangle $,
for which statistical test results for all experimentally tested laser
waists and powers are presented in Table. \ref{tab:Summary first-order click Z-stat}.
In contrast to the comparisons presented above for total counts and
multi-dimensional GCPs, added decoherence does little to improve statistical
test outputs, with all data sets deviating significantly from theory.

\begin{figure}
\begin{centering}
\includegraphics[width=0.95\columnwidth]{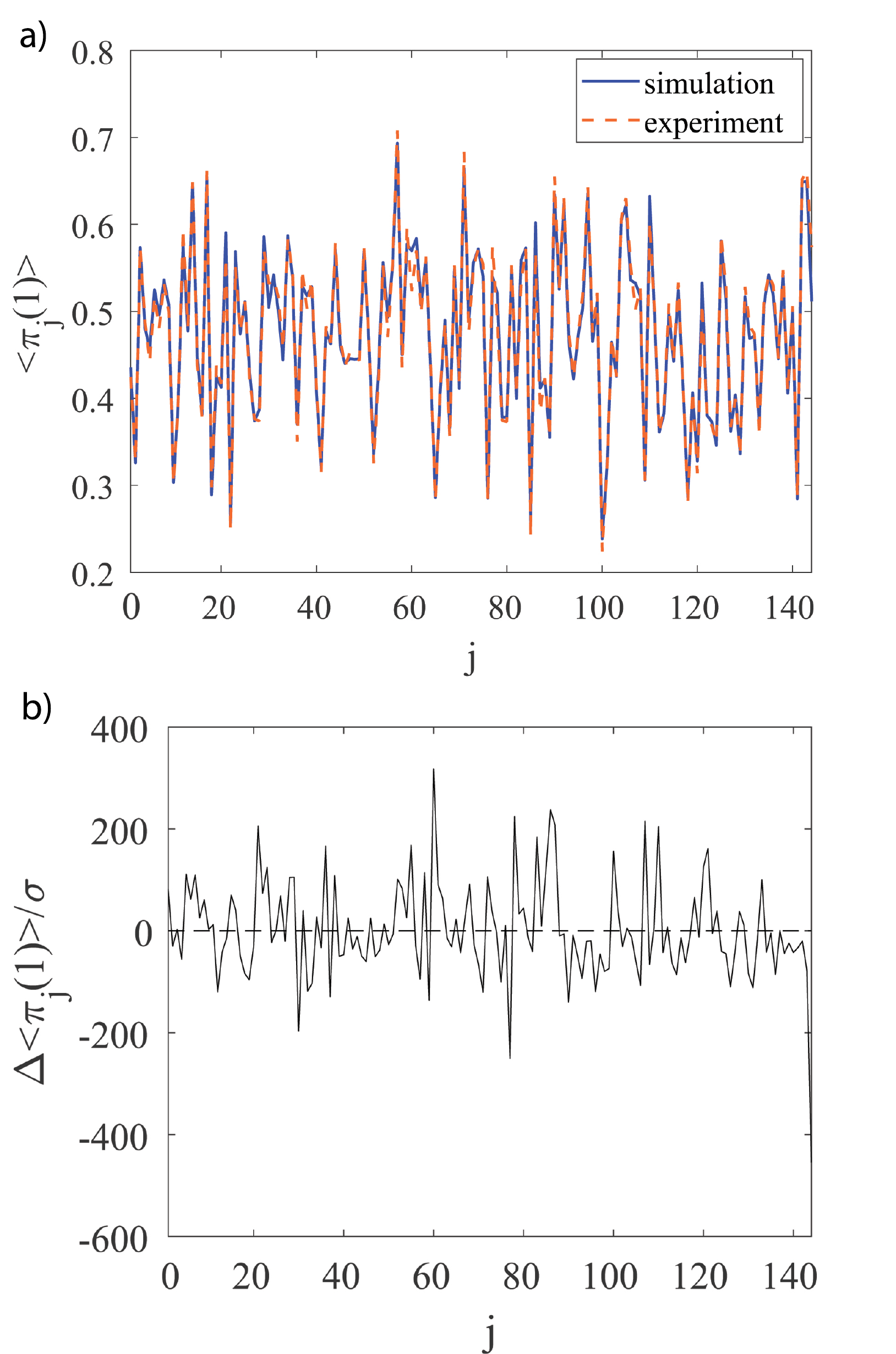}
\par\end{centering}
\caption{Comparison of theory versus experiment for all possible combinations
of first-order count probabilities, $\mathcal{G}_{\{j\}}^{(1)}(1)=\left\langle \hat{\pi}_{j}(1)\right\rangle $,
per channel $j$. Experimental samples are obtained from data set
$65\mu m$, $1.65W$ whilst simulations are performed using $E_{S}=1.44\times10^{7}$
ensembles and a small admixture of thermal decoherence. a) Comparison
plot with simulations represented by the solid blue and experimental
distributions plotted with orange dashed lines. b) Normalized difference
of $\Delta\left\langle \hat{\pi}_{j}(1)\right\rangle /\sigma_{m}$
versus $j$. \label{fig:First-order click comp}}
\end{figure}

Figure \ref{fig:First-order click comp} plots the first-order click
correlation for $65\mu m$, $1.65W$ where simulations contains a
small admixture of thermal decoherence. Although comparison plots
appear visually matching, graphed normalized differences tell a different
story (see Fig. (\ref{fig:First-order click comp}).b). This is reflected
in statistical testing, where $Z_{ET}\approx516$ and $Z_{EI}\approx541$. 

\begin{figure}
\begin{centering}
\includegraphics[width=0.95\columnwidth]{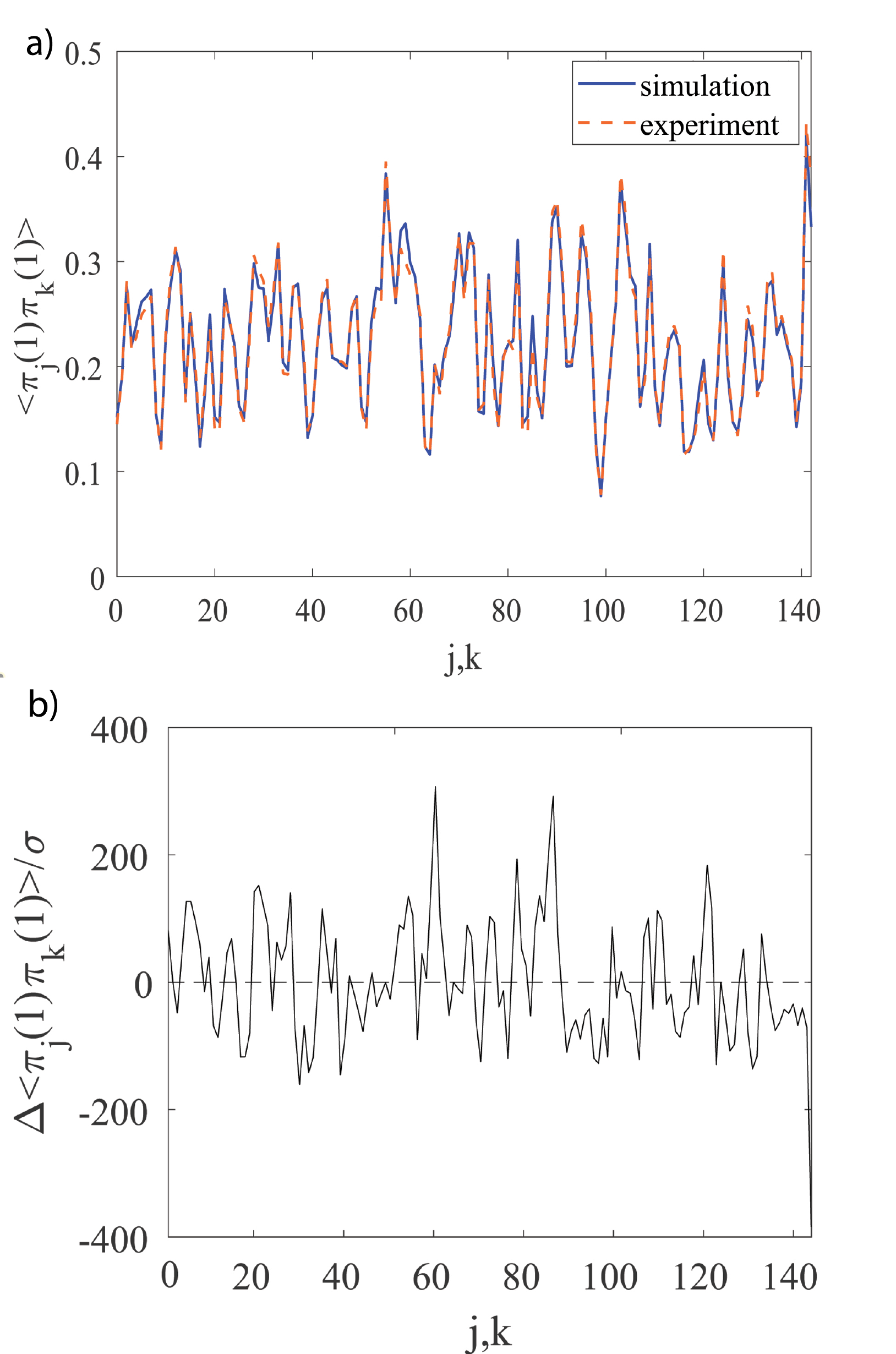}
\par\end{centering}
\caption{Theoretical and experimental comparison of a small sample of $143$
out of a possible $\binom{144}{2}=10296$ second-order click correlation
moments for modes $j,k$. Phase-space simulations use $E_{S}=1.44\times10^{7}$
ensembles whilst experimental samples are from $65\mu m$, $1.65W$,
with input decoherence corresponding to Table. \ref{tab:Total counts chi-square}.
a) Plotted comparisons of $\mathcal{G}_{\{j,k\}}^{(2)}(2)=\left\langle \hat{\pi}_{j}(1)\hat{\pi}_{k}(1)\right\rangle $
versus channels $j,k$ where simulations are represented by the solid
blue and experimental outputs are plotted with orange dashed lines.
b) Normalized difference of $\Delta\left\langle \hat{\pi}_{j}(1)\hat{\pi}_{k}(1)\right\rangle /\sigma_{m}$
versus $j,k$. \label{fig:Second-order click comp}}
\end{figure}

\begin{figure}
\begin{centering}
\includegraphics[width=0.95\columnwidth]{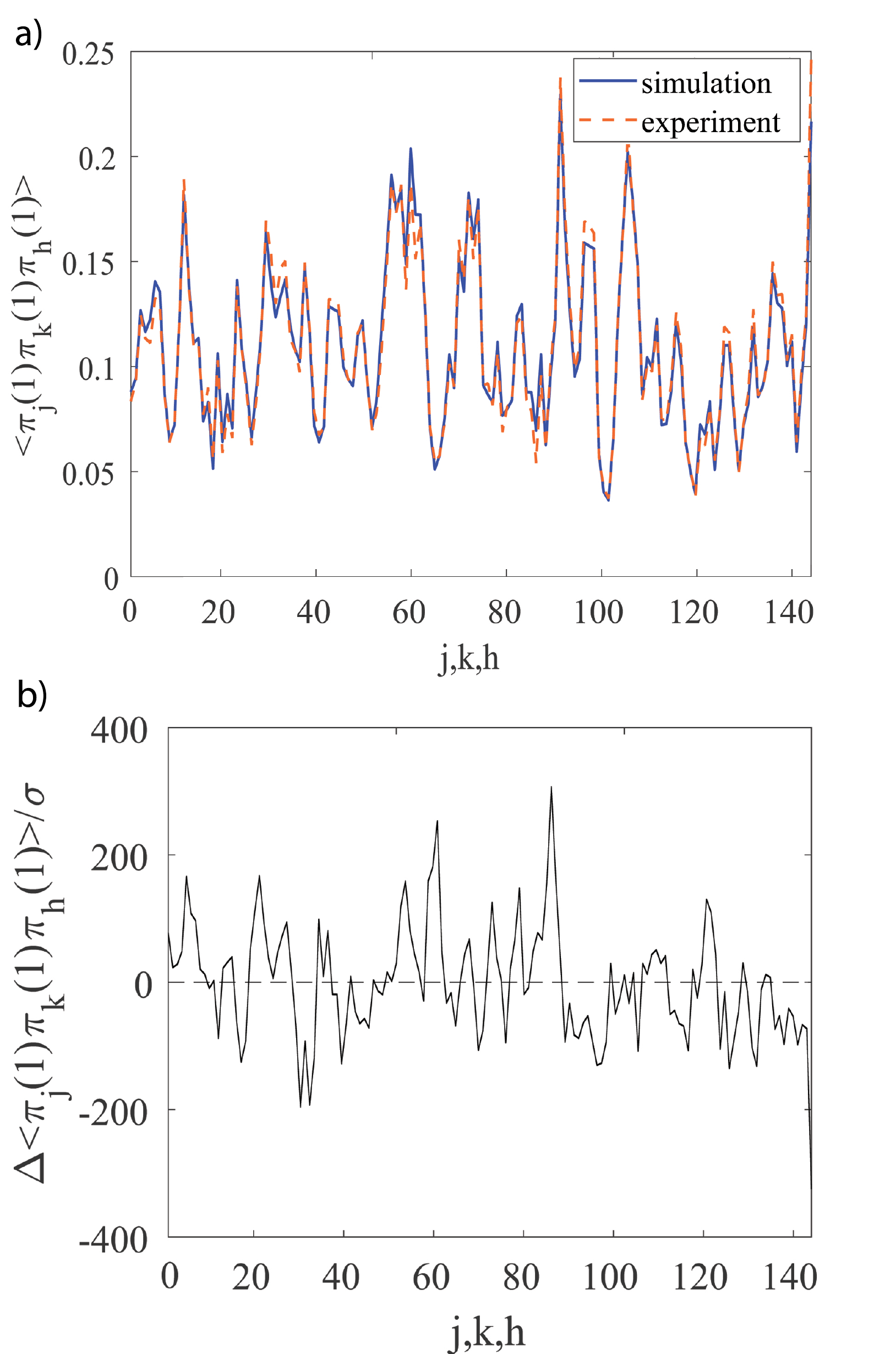}
\par\end{centering}
\caption{Errors in comparisons of theory and experiment for a subset of $142$
third-order click correlations, $\mathcal{G}_{\{j,k,h\}}^{(3)}(3)=\left\langle \hat{\pi}_{j}(1)\hat{\pi}_{k}(1)\hat{\pi}_{h}(1)\right\rangle $,
versus modes $j,k,h$. Statistical tests are performed using all $\binom{144}{3}=487344$
possible combinations of output modes. a) Graphical comparisons with
simulations denoted by the solid blue and obtained using $E_{S}=1.44\times10^{7}$
ensembles with added decoherence in the input stochastic samples,
whilst the orange dashed lines denotes experimental outputs from data
set $65\mu m$, $1.65W$. b) Normalized difference of $\Delta\left\langle \hat{\pi}_{j}(1)\hat{\pi}_{k}(1)\hat{\pi}_{h}(1)\right\rangle /\sigma_{m}$
versus $j,k,h$. \label{fig:Third-order click comp}}
\end{figure}

Comparisons of correlation moments of orders $n=2,3$ show these deviations
increase with order. We perform statistical tests for all possible
combinations of output modes for data from $65\mu m$, $1.65W$, where
the number of possible combinations of output modes follows $\binom{M}{n}=M!/(n!(M-n)!)$.
For graphical simplicity, only a small sample of these combinations
are plotted in Fig.(\ref{fig:Second-order click comp}), for $n=2$,
and Fig.(\ref{fig:Third-order click comp}) with $n=3$. 

As in the first-order case, although theoretical and experimental
distributions may generate visually similar outputs, $Z$-statistic
results of $Z_{ET}\approx4.3\times10^{3}$ and $Z_{ET}\approx2.7\times10^{4}$
for $n=2,3$, respectively show every increase in correlation order
sees an, approximately, order of magnitude increase in Z distances
that measure the statistical errors. 

An increase in statistical comparison errors with correlation order
is not surprising, because the quantity of data compared increases
with order. In general, large discrepancies may indicate systematic
errors, since outputs are far from their expected mean. Similar effects
are observed in the higher dimensional grouped counts, above.

These systematic errors could be caused by variations in transmission
or detector efficiency that are not accounted for in the measured
parameters, which could be included by fitting the output count data
by varying the transmission matrix $T$. However, at least $144$
output fitting parameters would be needed, making these comparisons
less meaningful.

\subsection{Summary}

In summary, experimental data does not agree with the ideal state
distributions within sampling error, measured by the Z-statistic.
This was found for all comparisons. Agreement is greatly improved
if the comparison is made with a thermalized quantum model. There
are lower discrepancies in total count data compared to mode-dependent
data. This suggests that the GBS model could be further refined with
better transmission parameter estimates.

\section{Classically generated photon counts\label{sec:Classical_fake}}

There is no known efficient classical method to generate the random
binary counts, if the input is a squeezed quantum state \citet{lund2014boson,quesadaQuadraticSpeedUpSimulating2022,quesada2020exact}.
However, classical states input into a linear bosonic network with
normally ordered detectors generate an output state which is efficiently
simulable using Glauber's diagonal P-representation \citep{qi2020regimes,RahimiKeshari:2016}.
This covers thermal, coherent \citep{Reid1986,Drummond2004_book,walls2008quantum},
and the recently proposed squashed states \citep{martinez-cifuentesClassicalModelsAre2022,jahangiriPointProcessesGaussian2020}. 

Although experimental GBS data is poorly modeled by thermal states
\citep{zhong2020quantum,zhongPhaseProgrammableGaussianBoson2021,villalonga2021efficient},
an analysis of experimental data from 100-mode GBS experiments showed
that this experiment was best fitted with a strongly thermalized input
\citep{drummondSimulatingComplexNetworks2022}. A thermal component
of $\epsilon=0.0932\pm0.0005$ was used to model this experiment,
which is approximately double the largest $\epsilon$ value presented
in Table \ref{tab:Total counts chi-square}. 

Consistent with this, a classical approximation to squeezed states,
called squashed states \citep{martinez-cifuentesClassicalModelsAre2022},
has been shown to agree with the $100$-mode experimental data at
least as well as pure state GBS predictions. This is not surprising,
as high thermalization levels produce Gaussian states which are nearly
classical.

The simplicity in computing output distributions of classical states
can be exploited to efficiently generate the corresponding binary
random numbers without the exponential time taken to (classically)
generate the ideal GBS distribution. This allows comparisons between
counts generated with a classical model and ideal GBS theory.

The number of probability distributions of interest can now be extended,
of which we are interested in the following four cases:
\begin{description}
\item [{(I)}] Ideal theoretical GBS squeezed state distributions
\item [{(E)}] Experimentally measured count distributions from data
\item [{(T)}] Thermalized, but quantum, best fit distributions
\item [{(C)}] Classical squashed state count distribution data
\end{description}
Using the notation in the Appendix, there are six relevant probability
distances that one can measure. The important ones are shown schematically
in Figs (\ref{fig:Ideal-picture}) and (\ref{fig:Real-picture}).
For an ideal GBS experiment, one expects that $Z_{EI}\approx1$, while
$Z_{CI}\gg1$. This would show that the experiment generates the expected
exponentially hard random count distributions, but the efficient classical
model is unable to generate the same data within sampling errors.

In real experiments, the inputs behave as though thermalized, possibly
from mode mismatch. Even this thermalized input can still have quantum
effects, provided one can show that $Z_{ET}\approx1$, while $Z_{CT}\gg1$
for at least one of the grouped count measures. To prove quantum advantage
in this practical but non-ideal case, one must show that this is true
for all possible efficient classical random algorithms.

When compared to experimental data from the $144$-mode network of
Ref.\citep{zhongPhaseProgrammableGaussianBoson2021}, low-order correlations
were found to be a very similar distance from classical squashed states
as the ideal GBS marginal probabilities \citep{martinez-cifuentesClassicalModelsAre2022}.
However, as explained in Section (\ref{subsec:Low-order-moments}),
low-order correlations are not an optimal measure of quantum versus
classical behavior.

Therefore, using the diagonal P-representation, which is a limit of
the generalized P-representation, we greatly extend the previous classical
count analysis in several extremely important directions: Firstly,
we compare experimental networks with simulated classical output distributions,
computing the distance measure $Z_{EC}$ for both low and high-order
correlations. This has the advantage of including sampling errors,
which weren't analyzed previously. 

Secondly, in order to compute the distance between the classical squashed
distribution and the theoretical ideal GBS, corresponding to the distance
measure $Z_{CI}$, we generate classically faked random binary patterns.
These patterns are generated using squeezing parameters and transmission
matrix corresponding to experimental networks $65\mu m,1.65W$ and
$65\mu m,0.15W$. Comparisons are also made between the classical
fakes and the best-fit thermalized model.

\subsection{Classical Gaussian boson sampling}

A classically simulable linear bosonic network is one in which $N\subset M$
classical states are input into a GBS where the input density operator
can be defined in terms of the diagonal P-representation as:

\begin{equation}
\hat{\rho}^{(\text{in)}}=\int P(\boldsymbol{\alpha})\left|\boldsymbol{\alpha}\right\rangle \left\langle \boldsymbol{\alpha}\right|\text{d}^{2}\boldsymbol{\alpha}.
\end{equation}

The diagonal-P distribution $P(\boldsymbol{\alpha})$ for classical
states is obtained from Eq.(\ref{eq:sigma_DP_distribution}) with
$\sigma=0$. As one simple example, input thermal states with variance
given by Eq.(\ref{eq:thermal_variance}), are defined by the distribution
\citep{walls2008quantum,gardiner2004quantum}

\begin{equation}
P(\boldsymbol{\alpha})=\prod_{j}\left(\frac{1}{\pi n_{j}}e^{-|\alpha_{j}|^{2}/n_{j}}\right).\label{eq:thermal_DP_distribution}
\end{equation}
Squashed states, with variance Eq.(\ref{eq:squashed_variance}), are
distributed in phase-space following \citep{jahangiriPointProcessesGaussian2020}

\begin{equation}
P(\bm{\alpha})=\prod_{j}\sqrt{\frac{1}{2\pi n_{j}}}e^{-\text{Re}\left(\alpha_{j}\right)^{2}/2n_{j}}\delta\left(\text{Im}\left(\alpha_{j}\right)\right).\label{eq:squashed_DP_distribution}
\end{equation}

Since both inputs are Gaussian, initial coherent amplitudes $\boldsymbol{\alpha},\boldsymbol{\alpha}^{*}$
are sampled efficiently from Eq.(\ref{eq:sigma_input_stochastic}),
where for thermal states one has 

\begin{equation}
\alpha_{j}=\sqrt{\frac{n_{j}}{2}}\left(w_{j}+iw_{j+M}\right),\label{eq:thermal_stochastic}
\end{equation}
whilst for squashed states

\begin{equation}
\alpha_{j}=\alpha_{j}^{*}=\sqrt{n_{j}}w_{j}.\label{eq:squashed_stochastic}
\end{equation}

As outlined in subsection \ref{subsec:Output-density-matrix} for
nonclassical states, output coherent amplitudes are obtained via the
transformation

\begin{align}
\boldsymbol{\alpha}' & =\boldsymbol{T}\boldsymbol{\alpha}\nonumber \\
\boldsymbol{\alpha}'^{*} & =(\boldsymbol{T}\boldsymbol{\alpha})^{*},\label{eq:DP_input_output}
\end{align}
since we are now in a classical phase-space, with corresponding output
density matrix 

\begin{equation}
\hat{\rho}^{(\text{out)}}=\int P(\boldsymbol{\alpha})\left|\boldsymbol{T\alpha}\right\rangle \left\langle \boldsymbol{T\alpha}\right|\text{d}^{2}\boldsymbol{\alpha}.\label{eq:output DP}
\end{equation}

In order to reproduce simulations performed in Ref. \citep{martinez-cifuentesClassicalModelsAre2022},
output coherent amplitudes are used to simulate GCPs. Following from
Eq.(\ref{eq:sigma moment equivalence}) for $\sigma=1$, GCPs, as
defined in Eq.(\ref{eq:grouped prob}), are now simulated over a classical
phase-space such that

\begin{equation}
\mathcal{G}_{\boldsymbol{S}}^{(n)}\left(\boldsymbol{m}\right)=\int P(\boldsymbol{\alpha})\left[\prod_{j=1}^{d}\left[\sum_{\sum c_{i}=m_{j}}\Pi_{S_{j}}\left(\boldsymbol{c}\right)\right]\right]\text{d}^{2}\bm{\alpha},\label{eq:DP_GCP}
\end{equation}
Here, $\Pi\left(\bm{c}\right)=\bigotimes_{j=1}^{M}\pi_{j}\left(c_{j}\right)$
is the phase-space observable for a binary pattern $\boldsymbol{c}$,
where $\pi_{j}$ has the same form as Eq.(\ref{eq:phase-space_click})
except now the output photon number is defined as $n'_{j}=|\alpha'_{j}|^{2}$. 

These are not the only efficient classical algorithms that one could
use to generate count data, but here we focus on the squashed state
case for the purpose of making definite comparisons.

\subsubsection{Generating classical fake bit patterns\label{subsec:Generating-classical-fake}}

In order to determine whether the experimental networks of Ref.\citep{zhongPhaseProgrammableGaussianBoson2021}
generate samples that are closer to the ideal GBS distribution than
a classical fake, as one would expect, we also use the coherent amplitudes
$\boldsymbol{\alpha}'$, $\boldsymbol{\alpha}'^{*}$ to generate fake
binary patterns. 

This is possible due to classical inputs generating an output distribution
that can be sampled efficiently \citep{qi2020regimes,villalonga2021efficient}.
For example, the first-order sampler of Ref.\citep{villalonga2021efficient}
efficiently generated binary patterns by approximating the output
of the bosonic network as a thermal state. Although such samplers
didn't implement phase-space methods, both methods produce similar
results.

Although this method can be used for any classical input state, we
focus on squashed states, since thermal states have been well tested
\citep{zhong2020quantum,zhongPhaseProgrammableGaussianBoson2021,villalonga2021efficient}. 

Initially, the process follows that outlined above; Stochastic amplitudes
for $N\subset M$ squashed inputs are generated using Eq.(\ref{eq:squashed_stochastic})
and are transformed to outputs following Eq.(\ref{eq:DP_input_output}).
The click probability of the $j$-th detector for each individual
stochastic trajectory is now computed as 

\begin{equation}
(\pi_{j}(1))^{(k)}=(1-e^{-n'_{j}})^{(k)},
\end{equation}
where $k\in E_{S}$ is a single trajectory in phase-space. 

This is straightforwardly extended to the multi-mode case as 

\begin{equation}
\boldsymbol{\pi}^{(k)}=[(\pi_{1}(1))^{(k)},(\pi_{2}(1))^{(k)},\dots,(\pi_{M}(1))^{(k)}],
\end{equation}
which is a vector of click probabilities for a single trajectory. 

Binary patterns can now be efficiently generated by randomly sampling
each bit independently using the Bernoulli distribution, which for
the $j$-th mode is defined as 

\begin{equation}
P_{j}^{(k)}(c_{j}^{(\text{fake)}})=(p_{j}^{(k)})^{c_{j}^{(\text{fake)}}}(1-p_{j}^{(k)})^{1-c_{j}^{(\text{fake)}}},
\end{equation}
where $p_{j}^{(k)}=(\pi_{j}(1))^{(k)}$ is the probability of generating
the fake click $c_{j}^{(\text{fake)}}=1$. Therefore, each trajectory
outputs the fake count vector 

\begin{equation}
(\boldsymbol{c}^{(\text{fake})})^{(k)}=[P_{1}^{(k)},P_{2}^{(k)},\dots,P_{M}^{(k)}].
\end{equation}

It is clear that the number of classical fake patterns generated scales
as $N_{F}=E_{S}$. For these to be useful for comparisons, they are
binned to generated grouped counts $m_{j}^{(\text{fake})}=\sum_{i=1}^{M}c_{i}^{(\text{fake})}$.
Following Eq.(\ref{eq:experiment_probability}), we define the GCP
of the $i$-th fake count bin as 

\begin{equation}
\mathcal{G}_{i}^{f}=\frac{m_{j_{i}}^{(\text{fake})}}{N_{F}}.\label{eq:fake_GCP}
\end{equation}

\subsection{Comparisons of classical GBS and experiment\label{subsec:Comp-classical-vs-exp}}

Analogous to Ref.\citep{martinez-cifuentesClassicalModelsAre2022},
we begin by comparing phase-space simulated GCPs of squashed state
inputs to experimentally binned patterns. Here, the distance between
experimental and classical GCPs are denoted by the subscript $EC$
(see Appendix for analytical definitions).

\begin{table*}
\begin{centering}
\begin{tabular}{ccccccc}
\toprule 
\multirow{2}{*}{Waist} & \multirow{2}{*}{Power} & \multicolumn{1}{c}{First-order moment} & \multicolumn{2}{c}{One-dimensional GCP} & \multicolumn{2}{c}{Two-dimensional GCP}\tabularnewline
\cmidrule{3-7} \cmidrule{4-7} \cmidrule{5-7} \cmidrule{6-7} \cmidrule{7-7} 
 &  & $Z_{EC}$ & $Z_{EC}$ & $k$ & $Z_{EC}$ & $k$\tabularnewline
\midrule
\midrule 
\multirow{2}{*}{$125\mu m$} & $1.412W$ & $318$ & $371$ & $49$ & $616$ & $663$\tabularnewline
\cmidrule{2-7} \cmidrule{3-7} \cmidrule{4-7} \cmidrule{5-7} \cmidrule{6-7} \cmidrule{7-7} 
 & $0.5W$ & $463$ & $489$ & $27$ & $735$ & $249$\tabularnewline
\midrule 
\multirow{5}{*}{$65\mu m$} & $1.65W$ & $536$ & $204$ & $83$ & $394$ & $1545$\tabularnewline
\cmidrule{2-7} \cmidrule{3-7} \cmidrule{4-7} \cmidrule{5-7} \cmidrule{6-7} \cmidrule{7-7} 
 & $1W$ & $397$ & $160$ & $72$ & $266$ & $1227$\tabularnewline
\cmidrule{2-7} \cmidrule{3-7} \cmidrule{4-7} \cmidrule{5-7} \cmidrule{6-7} \cmidrule{7-7} 
 & $0.6W$ & $281$ & $233$ & $54$ & $363$ & $782$\tabularnewline
\cmidrule{2-7} \cmidrule{3-7} \cmidrule{4-7} \cmidrule{5-7} \cmidrule{6-7} \cmidrule{7-7} 
 & $0.3W$ & $267$ & $399$ & $37$ & $634$ & $408$\tabularnewline
\cmidrule{2-7} \cmidrule{3-7} \cmidrule{4-7} \cmidrule{5-7} \cmidrule{6-7} \cmidrule{7-7} 
 & $0.15W$ & $234$ & $459$ & $24$ & $675$ & $204$\tabularnewline
\bottomrule
\end{tabular}
\par\end{centering}
\caption{$Z$-statistical test outputs of comparisons of all $144$ first-order
correlation moments, one and two-dimensional GCPs for all experimental
data sets. Phase-space simulations for squashed state inputs are performed
for $E_{S}=1.44\times10^{7}$ensembles for moment comparisons and
$E_{S}=1.2\times10^{6}$ for GCPs. In all cases, the subscript $EC$
denotes the distance between experimental and classical probabilities.
Analogous comparisons with nonclassical inputs are presented in Tables.
\ref{tab:Total counts chi-square}, \ref{tab:Summary_multi_dim_Z}
and \ref{tab:Summary first-order click Z-stat}. \label{tab:Summary_Z_exp_vs_classical}}
\end{table*}

\subsubsection{Low-order correlations}

Comparisons of first-order click moments are presented in Table. \ref{tab:Summary_Z_exp_vs_classical}
for every data set. When compared to the distance measures shown in
Table. \ref{tab:Summary first-order click Z-stat} for nonclassical
inputs, it is clear that for data set $65\mu m,1.65W$ one has $Z_{EC}\gtrsim Z_{EI}$,
as reported previously \citep{martinez-cifuentesClassicalModelsAre2022}. 

Increasing the correlation order to $n=2$ sees $Z$ values follow
a similar trend with $Z_{EC}\approx4.7\times10^{3}$ compared to nonclassical
inputs with $Z_{EI}\approx4.8\times10^{3}$ and $Z_{ET}\approx4.3\times10^{3}$.
Meanwhile, third-order moments output $Z_{EC}\approx3.1\times10^{4}$
whilst $Z_{EI}\approx3.1\times10^{4}$, $Z_{ET}\approx2.7\times10^{4}$
for pure and thermalized squeezed input comparison simulations. 

For all other data sets, bar $65\mu m,1W$ where $Z_{EI}\approx Z_{ET}\approx Z_{EC}$,
first-order moments generated from nonclassical inputs are closer
to the experimental marginals than classical inputs. This isn't surprising
for these data sets, as classical inputs into a linear network will
generate classical correlations, which are fundamentally different
from quantum correlations. As indicated by the large $Z$-statistic
values, one would expect a classical GBS to display further non-random
behavior in the click probabilities than an experiment containing
true quantum correlations. 

\subsubsection{Multi-dimensional binning}

Increasing the correlation order to $n=M$ and including all output
modes in the subset vector $\boldsymbol{S}$, comparisons of one and
two-dimensional GCPs are given in Table. \ref{tab:Summary_Z_exp_vs_classical}. 

Z-statistic outputs for all data sets other than $65\mu m,1.65W$
satisfy $Z_{EC}>Z_{EI}$ for both total count and two-dimensional
GCPs. Therefore, for these data sets, output distributions of simulated
squashed states poorly model experimental distributions, with simulated
pure squeezed states giving closer agreement with experiment. 

This is not the case for data set $65\mu m,1.65W$, with comparisons
outputting $Z_{EC}\approx204$ and $Z_{EC}\approx394$ for simulations
of one and two-dimensional GCPs, respectively. When compared to simulations
of the ideal GBS, we see that $Z_{EI}\gtrsim Z_{EC}$. For both low-order
moments and multi-dimensional GCPs, simulated ideal and squashed classical
distribution are approximately the same distance to experimentally
binned patterns from data set $65\mu m,1.65W$. 

Although these results agree with previous comparisons \citep{martinez-cifuentesClassicalModelsAre2022},
experimental samples are just as poorly modeled by squashed inputs
as by the ideal GBS. We emphasize that the best agreement between
theory and experiment for all possible data sets arises from nonclassical,
partially thermalized squeezed state inputs, which output the lowest
$\chi^{2}$ and $Z$-statistic values for all tested GCP dimensions. 

\subsection{Comparisons of classical and quantum output distributions}

Using the method outlined in subsection \ref{subsec:Generating-classical-fake},
we generate $N_{F}=4\times10^{7}$ classical binary patterns from
$E_{S}=4\times10^{7}$ initial stochastic samples of input squashed
states. 

Two sets of classically faked patterns are generated using squeezing
parameters and transmission matrix $\boldsymbol{T}$ from data sets
$65\mu m,1.65W$ and $65\mu m$, $0.15W$. These patterns are binned
following Eq.(\ref{eq:fake_GCP}) and compared to the ideal GBS distribution,
which is simulated using $E_{S}=1.2\times10^{6}$ ensembles for total
counts and two-dimensional GCPs, whilst marginal probabilities are
simulated using $E_{S}=1.44\times10^{7}$. 

Two distance measures are of interest in this section; The distance
between faked and ideal GCPs, $Z_{IC}$, and the distance between
faked and thermalized GCPs, $Z_{CT}$. The first is the usual measure
used to dispute claims of quantum advantage for classical inputs \citep{zhong2020quantum,zhongPhaseProgrammableGaussianBoson2021,madsenQuantumComputationalAdvantage2022,villalonga2021efficient},
whilst the second allows one to determine whether the more realistic
nonclassical thermalized distributions can be faked via classical
inputs. 

\begin{table*}
\begin{centering}
\begin{tabular}{ccccccccc}
\toprule 
 & \multicolumn{4}{c}{$65\mu m$, $1.65W$} & \multicolumn{4}{c}{$65\mu m$, $0.15W$}\tabularnewline
\midrule 
 & $Z_{EI}$ & $Z_{ET}$ & $Z_{CI}$ & $Z_{CT}$ & $Z_{EI}$ & $Z_{ET}$ & $Z_{CI}$ & $Z_{CT}$\tabularnewline
\midrule
\midrule 
$\left\langle \hat{\pi}_{j}(1)\right\rangle $ & $541$ & $516$ & $98$ & $253$ & $198$ & $196$ & $108$ & $137$\tabularnewline
\midrule 
$\left\langle \hat{\pi}_{j}(1)\hat{\pi}_{k}(1)\right\rangle $ & $4.8\times10^{3}$ & $4.3\times10^{3}$ & $856$ & $2.7\times10^{3}$ & $2.2\times10^{3}$ & $2.2\times10^{3}$ & $1.5\times10^{3}$ & $1.4\times10^{3}$\tabularnewline
\midrule 
$\mathcal{G}_{144}^{(144)}(m)$ & $221$ & $23$ & $122$ & $198$ & $49$ & $0.7$ & $300$ & $290$\tabularnewline
\midrule 
$\mathcal{G}_{72,72}^{(144)}(m_{1},m_{2})$ & $422$ & $185$ & $199$ & $350$ & $76$ & $40$ & $468$ & $453$\tabularnewline
\bottomrule
\end{tabular}
\par\end{centering}
\caption{Summary of $Z$-statistic results for comparisons of squashed state
fake binary patterns with simulated pure and thermalized squeezed
states. $N_{F}=4\times10^{7}$ fake patterns are generated and binned
with squeezing parameters and transmission matrix $\boldsymbol{T}$
from experimental data sets $65\mu m$, $1.65W$ and $65\mu m$, $0.15W$,
while phase-space simulations are performed for $E_{S}=1.2\times10^{6}$.
The important distance measures in this table is the distance between
faked GCPs and output distributions of the ideal, $Z_{CI}$, and thermalized,
$Z_{CT}$, GBS instances. For ease of reference, comparisons presented
throughout the paper of experiment with the ideal, $Z_{EI}$, and
thermalized, $Z_{ET}$, distributions are also presented. \label{tab:Summary_class_fake} }
\end{table*}

\subsubsection{Classically faked marginal probabilities}

Since the diagonal P-representation only contains diagonal coherent
state amplitudes, classical states sent through a bosonic network
will only generate classical correlations and one would expect some
degree of non-randomness in the photon counts to occur. This made
clearer by comparing $Z$ values, which are presented in Table. \ref{tab:Summary_class_fake}
for first and second-order correlations. Expectedly, classically faked
first and second-order correlation moments are far from the ideal
for both tested data sets, with $Z_{CI}\approx98$ and $Z_{CI}\approx108$
for patterns generated using $\boldsymbol{r}$ and $\boldsymbol{T}$
from $65\mu m$, $1.65W$ and $65\mu m$, $0.15W$, respectively.

\begin{figure}
\begin{centering}
\includegraphics[width=0.95\columnwidth]{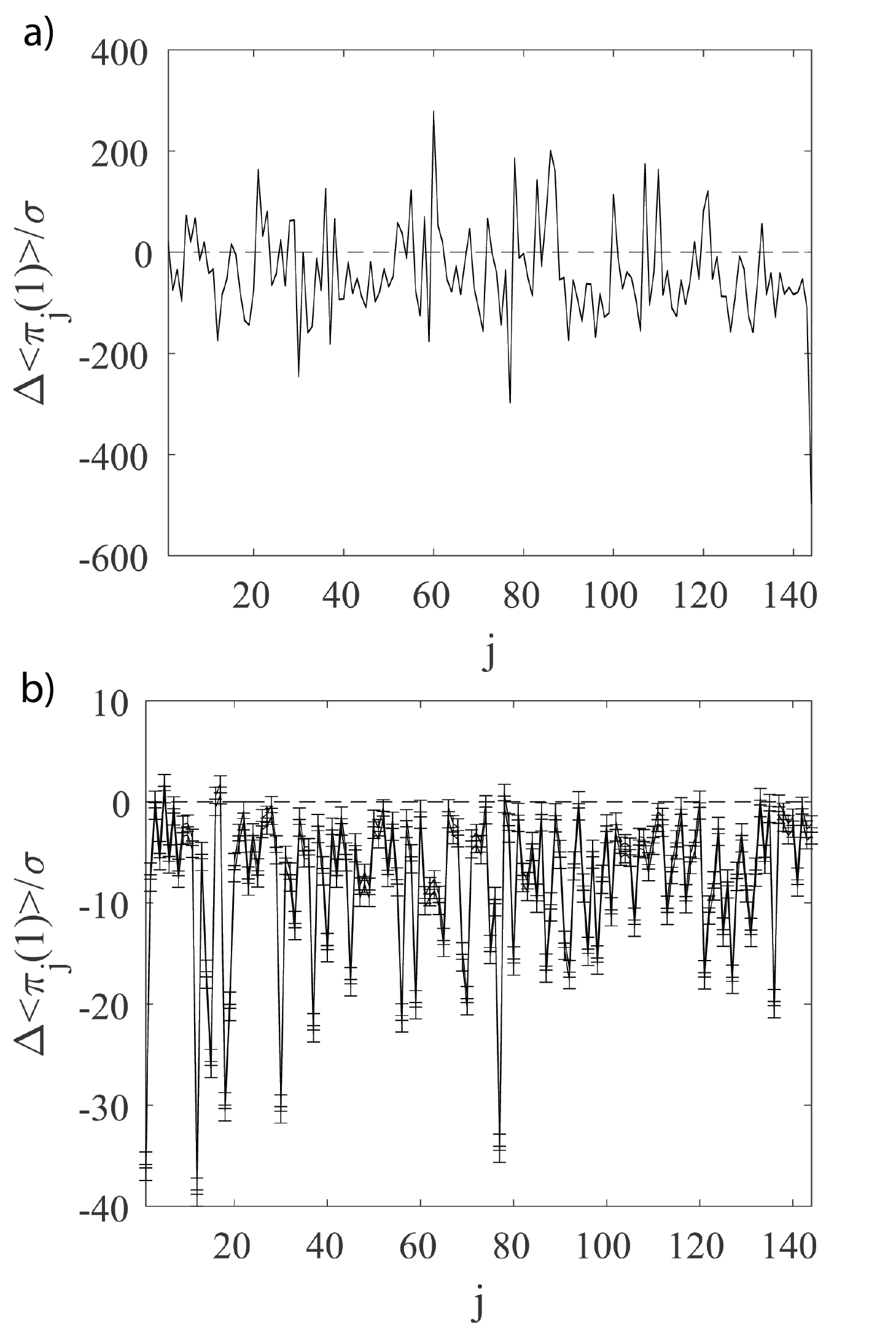}
\par\end{centering}
\caption{Normalized difference of first-order click correlation moments, $\left\langle \hat{\pi}_{j}(1)\right\rangle $,
simulated in phase-space using $E_{S}=1.44\times10^{7}$ ensembles
with pure squeezed state inputs which are compared to a) experimental
first-order moments from data set $65\mu m$, $1.65W$ and b) moments
generated from classical squashed state faked patterns generated using
squeezing parameters and transmission matrix from data set $65\mu m$,
$1.65W$. \label{fig:comp_click_norm_diff}}

\end{figure}

Despite this, when compared to experimental marginals, the classical
fakes for both data sets are closer to the ideal marginals than the
experiment. This is particularly noticeable for fakes obtained from
$65\mu m$, $1.65W$, which sees $Z_{EI}\approx5.5Z_{CI}$ for all
tested correlation orders. A graphical representation is presented
in Fig.\ref{fig:comp_click_norm_diff}, which compares the normalized
difference of experimental first-order marginals (Fig.\ref{fig:comp_click_norm_diff}a))
and squashed fakes (Fig.\ref{fig:comp_click_norm_diff}b)). 

Simulated thermalized squeezed inputs, with fitting parameters presented
in Table. \ref{tab:Total counts chi-square} for each data set, show
that the classical squashed fakes are also closer to the thermalized
marginals than the experiment. As with the ideal case, fakes generated
using $65\mu m$, $1.65W$ data set parameters are closest to the
thermalized first-order marginals, with experiments being $Z_{ET}-Z_{CT}=263\sigma$
further away from the simulated distribution than the squashed fakes. 

\subsubsection{Classically faked grouped count probabilities}

Although marginal probabilities allow for a direct comparison of specific
correlations, they are only part of the picture, and a more thorough
analysis is obtained when comparing multi-dimensional GCPs. Since
simulations of nonclassical states in phase-space reproduce high-order
quantum correlations generated in the network, one expects experimental
data to beat classical fakes, which only contain classical correlations.

As outlined in section \ref{sec:Comp_theory_exp}, experimental patterns
from data set $65\mu m$, $0.15W$ are closer to the pure and thermalized
squeezed state output distributions than any other data set. Therefore,
we first compare classical squashed fakes generated using squeezing
parameters and transmission matrix from this data set to determine
whether the thermalized and ideal output distributions can be efficiently
faked. 

Comparisons of classical fakes with the ideal output for both one
and two-dimensional GCPs shows that the classical fake is much further
from ideal than the experiment now that higher order correlations
are simulated. As seen in Table. \ref{tab:Summary_class_fake}, Z-statistic
tests output $Z_{CI}\approx300$ and $Z_{CI}\approx468$ for one and
two-dimensional GCPs, whilst experiments output a probability distance
of $Z_{EI}\approx40$ and $Z_{EI}\approx76$ for the same observables. 

Simulations of thermalized squeezed inputs indicate binned faked patterns
are even further from the expected distribution, with total counts
distances $Z_{CI}\approx290$ being well over two orders of magnitude
larger than the corresponding experimental outputs of $Z\approx0.7$.
Two-dimensional binning show differences with theory on increase for
comparisons with classical fakes. Clearly, experimental binary patterns
for data set $65\mu m$, $0.15W$ are much closer to both ideal and
thermalized distribution once higher-order correlations are considered
than their classically faked counterparts. 

A different story arises for classical fakes generated using $\boldsymbol{r}$
and $\boldsymbol{T}$ from experimental data set $65\mu m$, $1.65W$.
For all GCP dimensions tested, squashed fakes output smaller $Z$
values, and hence smaller $\chi^{2}/k$, for comparisons with simulated
pure squeezed state inputs than their experimental counterparts. Apart
from using a different classical count generator, this classical advantage
was found in a previous investigation of the same dataset \citet{villalonga2021efficient}.

\begin{figure}
\begin{centering}
\includegraphics[width=0.95\columnwidth]{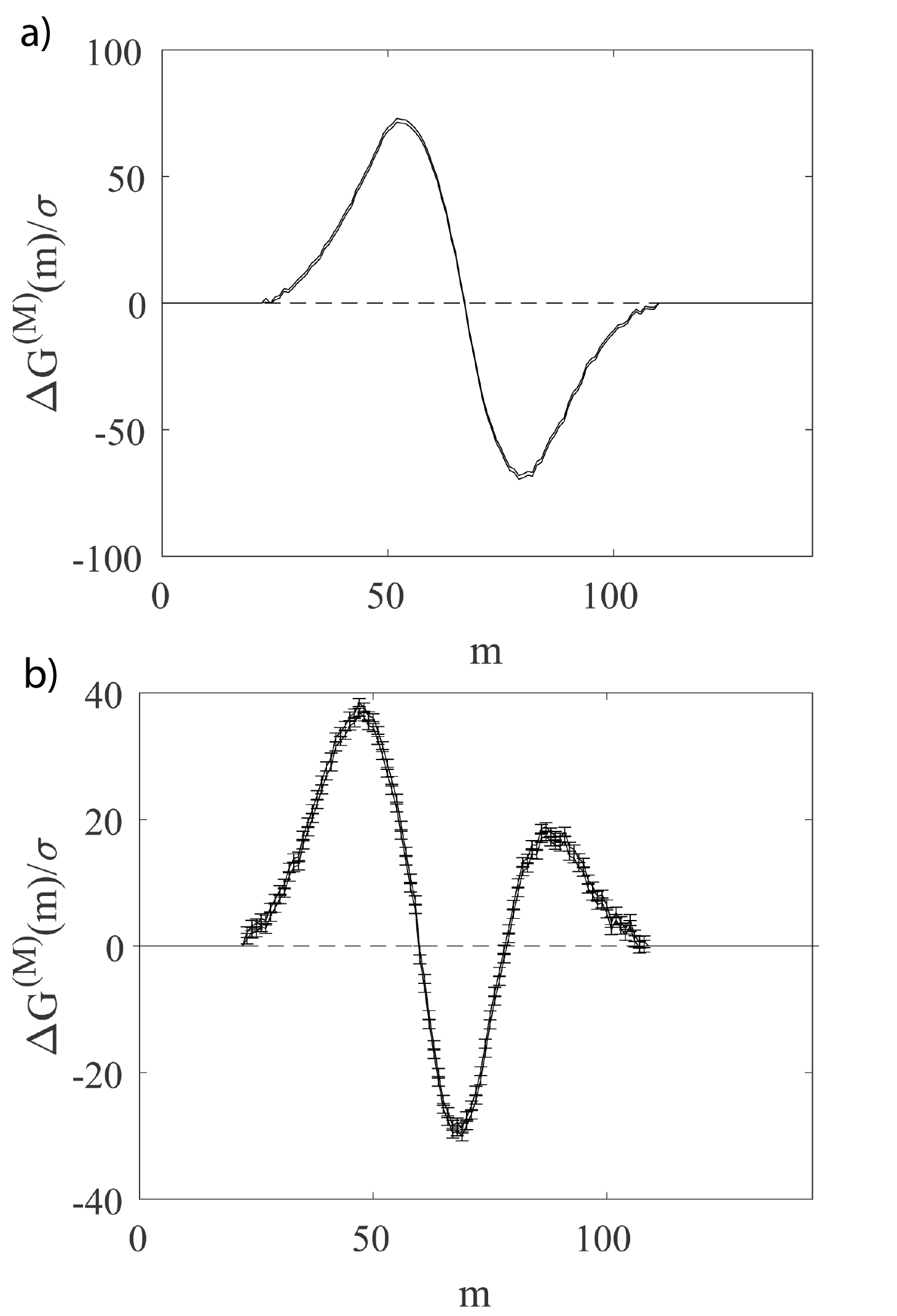}
\par\end{centering}
\caption{Total count, $\mathcal{G}_{144}^{(144)}(m)$, normalized difference
generated from comparisons of simulated pure squeezed states and a)
experimental data from $65\mu m$, $1.65W$ and b) fake binary patterns
generated using squeezing and transmission matrix from the same data
set.\label{fig:comp_totalc_norm_diff} }
\end{figure}

Although Z values still indicate a large degree of non-randomness
in present in the faked counts, differences are approximately half
that of the experiment (see Table. \ref{tab:Summary_class_fake})
which is made clear in comparisons of total count, where $Z_{EI}\approx221$
and $Z_{CI}\approx122$. This can also be seen in Fig. \ref{fig:comp_totalc_norm_diff}
which compares the normalized difference of total counts from experiment
and classical fakes. 

Increasing the dimension further to $d=4$ sees the $Z$-statistic
improve for comparisons with squashed fakes, giving $Z_{CI}\approx122$.
To make comparisons with experimental and fake patterns accurate,
we generate an identical number of samples. As described above, experimental
sampling errors become significant at $d=4$ due to the large number
of bins with too few photons per bin. Therefore, the decrease in Z
value of the four-dimensional faked GCP is likely due to an increase
in the sampling error as opposed to an improved agreement with theory. 

Despite the better agreement between ideal and classical fakes for
this data set, distributions corresponding to thermalized squeezed
inputs show better agreement with experiment than classical fakes.
Total count comparisons Z value of $Z_{ET}\approx23$ and $Z_{CT}\approx198$
show significant differences in probability measures, which continue
when the dimension is increased to $d=2$, with faked Z values being
$165\sigma$ further from the simulated thermalized distribution than
the experiment.

Increasing the dimension to $d=4$ further shows that classical fakes
poorly replicate thermalized probabilities, with $Z_{ET}\approx200$
and $Z_{CT}\approx342$, although both experiment and classical fake
$Z$ values indicate photon counts display significant errors once
higher order correlations are simulated. This result is expected for
classical inputs, but regardless of input state, the current generation
of experimental linear networks with saturating detectors, for which
computational advantage has been claimed, still show large deviations
from theoretical predictions for most of the outputs.

Possible reasons for differences include parameter errors, network
fluctuations and nonlinearities. These results highlight the need
for a more detailed model of losses and decoherence to fully analyze
current linear photonic networks. Despite this, it is encouraging
that some low-power total count data gives evidence of computational
advantage.

\subsection{Comparison summary}

In some cases, classical count simulations performed much better than
experimental counts in generating the ideal state distributions. For
computational target distributions of a best fit thermalized state,
the experiment performs better for the high order correlations. For
one low power experiment, we find that the experimental total count
distribution also agrees with thermalized theory to within the experimental
sampling error, as illustrated in Fig (\ref{fig:Real-picture}).

\section{Outlook}

Statistical testing of both low and high order correlations is essential
for the validation and performance analysis of any large-scale quantum
technology. These can detect departures from randomness in experimental
photon count data, show the presence of systematic errors and allow
one to differentiate experimental data from data spoofed via classical
algorithms. 

Error-free operation is crucial for any computer, including the present
application of linear networks as quantum random number generators
\citep{zhong2020quantum}. These tests allow one to determine the
extent of decoherence present in experiments, which can cause networks
to become classically simulable \citep{qi2020regimes}, leading to
improved designs in future. 

Using the efficient positive-P phase-space method introduced in \citep{drummondSimulatingComplexNetworks2022},
we implement correlation tests of all orders. These include higher
dimensional binning algorithms with a randomized test generator, giving
a dynamic verification tool that can both validate outputs and potentially
prevent faking. We show that, although the current generation of boson
sampling quantum computers have outputs that significantly differ
from ideal behavior, thermalized yet nonclassical squeezed state inputs
generate distributions with the smallest distance to experimental
distributions for all data sets tested. However, significant differences
persist even when this simple model of modified transmission and partial
decoherence is included.

Although errors are larger for inhomogeneous tests like single-channel
output counts, this could be caused by estimation errors in the transmission
parameters, which are removable with better values. These are less
significant than the total count distributions, which are sensitive
to a quantum input state. One case with low input powers gave a total
count distribution agreeing exactly with the partly thermalized total
count distribution.

A strategy of generating ``fake'' counts from squashed state inputs
is also implemented, and shown to have output distributions closer
to the ideal than the experiment, for data sets with large laser power.
However, at low power, and by extension low numbers of detected counts,
the experiment beats the classical squashed fakes, although still
with significant errors. Comparisons with a partly thermalized Gaussian
model and high order correlations showed that the quantum experiment
generated output statistics much closer to this slightly decoherent
model than the classical counts.

In summary, there is evidence of computational advantage, but not
for the original pure squeezed state input model. Instead, it is for
a targeted distribution that is more decoherent than the ideal Gaussian
boson sampling input. This is in general agreement with detailed analyses
of an earlier experiment \citep{drummondSimulatingComplexNetworks2022,delliosSimulatingMacroscopicQuantum2022a},
where the decoherence was so large that classical models gave similar
output statistics \citep{martinez-cifuentesClassicalModelsAre2022}.

Although these results demonstrate possible computational advantage,
they also show that experimental decoherence and other imperfections
cause a departure from the ideal model. We therefore suggest that
scalable simulators or other methods are essential to the verification
and benchmarking of a wide range of large-scale quantum technologies.
\begin{acknowledgments}
This work was partly performed on the OzSTAR national facility at
Swinburne University of Technology. OzSTAR is funded by Swinburne
University of Technology and the National Collaborative Research Infrastructure
Strategy (NCRIS). This research was funded through grants from NTT
Phi Laboratories and the Australian Research Council Discovery Program.
\end{acknowledgments}

\section*{Appendix A: Chi-squared tests and sampling errors}

The main statistical test used to quantify differences between phase-space
simulations of GCPs and experimental outputs is the chi-square test
\citep{pearson1900x}. Chi-square tests are a powerful statistical
test commonly used to determine whether observed probabilities obtained
from independent samples correspond to the predicted distribution
of the system being tested \citep{Rukhin2010,knuth2014art}.

Let $N_{E}$ denote the number of independent experimental observations.
These are classified into $k$ classes denoting all the possible outcomes
one can observe. Each class has a theoretical probability of $P_{i}$
with $i=1,2,\dots k$, where the expected number of observations of
the $i$-th class is $N_{E}P_{i}$, whilst the actual number of observations
from an experiment is $x_{i}$. 

The standard chi-square statistic is defined as \citep{knuth2014art}:

\begin{equation}
\chi^{2}=\sum_{i=1}^{k}\frac{\left(N_{E}P_{i}-x_{i}\right)^{2}}{N_{E}P_{i}},\label{eq:standard chi-square}
\end{equation}
which can be rewritten in terms of the estimated experimental probability
$P_{i}^{e}=x_{i}/N_{E}$ and variance $\sigma_{i}^{2}=P_{i}/N_{E}$. 

In terms of GCPs, we define the grouped count of the $i$-th class
as $m_{j_{i}}$ with each class representing a detector count bin.
Using the shorthand notation $\mathcal{G}_{i}$ to denote the true
theoretical GCP of the $i$-th class, the experimental GCP is obtained
using $\mathcal{G}_{i}^{e}=m_{j_{i}}/N_{E}$ for $N_{E}$ experimental
samples. 

Since both experimental and theoretical probabilities are obtained
via sampling, a modified version of Eq.(\ref{eq:standard chi-square})
is required \citep{drummondSimulatingComplexNetworks2022}:

\begin{equation}
\chi^{2}=\sum_{i=1}^{k}\frac{\left(\bar{\mathcal{G}}_{i}-\mathcal{G}_{i}^{e}\right)^{2}}{\sigma_{i}^{2}},\label{eq:grouped count chi-square}
\end{equation}
where $\mathcal{\bar{G}}_{i}$ is the phase-space simulated ensemble
mean with $\mathcal{G}_{i}=\lim_{E_{S}\rightarrow\infty}\mathcal{\bar{G}}_{i}$
and we define
\begin{equation}
\sigma_{i}^{2}=\sigma_{T,i}^{2}+\sigma_{E,i}^{2}
\end{equation}
as the sum of theoretical and experimental sampling errors. This ensures
that both the distribution variances are included. For an ideal case,
this combination is the theoretical variance of the difference between
probabilities $\Delta_{i}=\bar{\mathcal{G}}_{i}-\mathcal{G}_{i}^{e}$,
which is called the difference error. 

Due to Poissonian fluctuations, experimental sampling errors are estimated
as $\sigma_{E,i}\approx\sqrt{\mathcal{G}_{i}/N_{E}}$. These comparisons
include both experimental and theoretical variances for best accuracy.

Although the output errors follow a chi-square distribution, the input
probabilities in the $\chi^{2}$ sum are required to be both Gaussian,
or approximately Gaussian as is the usual case, and independent. Small
count numbers per bin means the experimental probability of that bin
can no longer be approximated as a Gaussian. Therefore, chi-square
tests are only performed for $k$ valid bins, which we define as those
having $m_{j_{i}}>10$ \citep{Rukhin2010,knuth2014art}.

\section*{Appendix B: Z-statistic tests}

Even though comparisons of measured probabilities with theory are
essential to validation tests, Z-statistics have not been used previously
to analyze GBS data. Here we give a brief account of this method.

For grouped data, with high enough counts per bin, the expected distributions
of measured counts are approximately Gaussian as described above.
Hence, for probabilities estimated from experimental measurements,
output results are expected to satisfy $\chi^{2}/k-1\approx0$. The
exact value of $\chi^{2}/k$ is then an important indicator that experimental
distributions have acceptable errors. 

We note that since there are some theoretical sampling errors, we
include these error-bars as well. However for experimental data currently
available it is not difficult to obtain enough theoretical samples
such that this is not an important issue. 

Because the experimental data, even after binning, has large numbers
of different probabilities, one finds that $k\gg1$. Hence tabulated
chi-square distributions for small $k$ are not useful.

When $k\rightarrow\infty$, the chi-square distribution $\chi^{2}\rightarrow\mathcal{N}(\mu,\sigma^{2})$
with mean $\mu=k$ and variance $\sigma^{2}=2k$, via the central
limit theorem \citep{wilsonDistributionChiSquare1931,johnsonContinuousUnivariateDistributions1970}.
This convergence is slow due to the skewness of the chi-square distribution
\citep{johnsonContinuousUnivariateDistributions1970}. Fortunately,
an accurate and fast convergence is achieved using the Wilson-Hilferty
(WH) transformation \citep{wilsonDistributionChiSquare1931}, where
the transformed observable $\left(\chi^{2}/k\right)^{1/3}\rightarrow\mathcal{N}(\mu,\sigma^{2})$
for $k\geq10$ with $\mu=1-\sigma^{2}$ and $\sigma^{2}=2/(9k)$ \citep{wilsonDistributionChiSquare1931,johnsonContinuousUnivariateDistributions1970}.

If the chi-square distribution can be approximated as a Gaussian,
we can then perform the Z-statistic, or Z-score, test which has the
general definition $Z=(X-\mu)/\sigma$ \citep{freundStatisticalMethods2003,Rukhin2010},
where $X$ is the test statistic. The Z-statistic determines how many
standard deviations a test statistic is from its normally distributed
expected mean. 

In terms of the WH transformed chi-square statistic $X=\left(\chi^{2}/k\right)^{1/3}$,
we define an approximate Z-statistic as:

\begin{equation}
Z=\frac{\left(\chi^{2}/k\right)^{1/3}-\left(1-2/(9k)\right)}{\sqrt{2/(9k)}}.
\end{equation}
The Z-statistic is used to determine the probability of obtaining
the observed chi-square output. A result of $Z>6$ indicates the test
statistic has an extremely small probability of being observed. 

In GBS with saturating detectors, if detector outcomes are random
and independent, one would expect $\left\langle \hat{\pi}_{j}(1)\right\rangle \approx\left\langle \hat{\pi}_{j}(0)\right\rangle $.
Therefore, the Z-statistic quantifies whether the observed chi-square
distribution corresponds to randomly distributed photon counts. If
$Z>6$, one can hypothesize that count patterns are displaying non-random
behavior. An example of such non-randomness would be experimental
probabilities where $\left\langle \hat{\pi}_{j}(1)\right\rangle \gg\left\langle \hat{\pi}_{j}(0)\right\rangle $.
Physically, this may correspond to errors occurring either in the
network, generation of input states or detectors themselves, e.g.
detector dark counts. 

However, the power of the Z-statistic becomes clearer when used to
test comparisons with random permutations of binary patterns. If each
permutation repeatedly produces chi-square outputs satisfying $\chi^{2}/k\gg1$,
then Z-statistic outputs in turn will satisfy $Z>6$. Therefore, the
output $\chi^{2}/k$ values would be highly unlikely to be observed
once, when compared to the mean of the resulting WH chi-square distribution,
let alone multiple times. This result would indicate systematic errors
may be present in the tested network causing events with small probabilities
to be continuously observed. 

\section*{Appendix C: Distance measure notation}

Comparisons of GCPs for multiple input states, both classical and
nonclassical, with experimental and classically faked binary patterns
are presented throughout this paper. Statistical tests of these comparisons
result in multiple probability distances being compared. Here, we
define each distance measure analytically relative to an estimated
count GCP and a theoretical GCP. 

As outlined above, the theoretical GCP observable is the phase-space
ensemble mean 

\begin{equation}
\bar{\mathcal{G}}_{i}=\frac{1}{E_{S}}\sum_{k=1}^{E_{S}}(\mathcal{G}_{i,\sigma})^{(k)},
\end{equation}
where $k$ denotes a single trajectory in the entire phase-space and
$\mathcal{G}_{i,\sigma}$ is the $\sigma$-ordered stochastic GCP. 

Two phase-space methods are used to simulate GCPs: The normally ordered
positive-P representation, allowing one to simulate pure and thermalized
squeezed states, and the diagonal P-representation, which is only
valid for classical input states. Ensemble mean GCPs of the $i$-th
class for simulations of pure squeezed states, the ideal output distribution,
are denoted as $\bar{\mathcal{G}}_{i,I}$, whilst outputs assuming
thermalized and classical input states are $\bar{\mathcal{G}}_{i,T}$
and $\bar{\mathcal{G}}_{i,C}$, respectively. 

The normalized distance between experimental GCPs estimated from experimental
counts, $\mathcal{G}_{i,E}$, and phase-space simulated theoretical
GCPs is computed using a chi-square statistical test which, for generality,
is now defined as 

\begin{equation}
\chi_{ES}^{2}=\sum_{i=1}^{k}\frac{\left(\mathcal{G}_{i,E}-\bar{\mathcal{G}}_{i,S}\right)^{2}}{\sigma_{i}^{2}},
\end{equation}
where $\bar{\mathcal{G}}_{i,S}$ is a stochastic ensemble mean for
any general input state, $S$, and $\sigma_{i}^{2}$ is the variance
estimate. Comparisons with classically generated (ie, fake) counts,
$\mathcal{G}_{i,C}$, give 

\begin{equation}
\chi_{CS}^{2}=\sum_{i=1}^{k}\frac{\left(\mathcal{G}_{i,C}-\bar{\mathcal{G}}_{i,S}\right)^{2}}{\sigma_{i}^{2}}.
\end{equation}

Each chi-square output is used to compute the $Z$-statistic test
as defined above. Therefore, $Z_{ES}$ and $Z_{CS}$ are the corresponding
$Z$ values for each probability difference measure. When classically
generated fake counts are compared with a classical phase-space moment,
they ought to agree within sampling error.

This is shown by $Z_{CC}$ values for two simulation tests presented
in the table below. Both tests simulate $N=50$ squashed inputs into
the $M=144$ mode experimental network. The first test simulates this
classical GBS using squeezing parameters and transmission matrix from
data set $65\mu m,0.15W$, whilst the second corresponds to $65\mu m,1.65W$.
In both cases, the transmission coefficient is $t=1$. 

Phase-space simulations for squashed states are compared to the classical
count simulations used for comparisons in the main text, which are
generated with the same squeezing and transmission matrix as used
in the main simulations. In all these test cases, $Z_{CC}\approx1$,
showing that the distribution moments of the two methods agree within
sampling error. 

\begin{table}
\begin{centering}
\begin{tabular}{ccc}
\toprule 
 & $Z_{CC}^{(0.15W)}$ & $Z_{CC}^{(1.65W)}$\tabularnewline
\midrule
\midrule 
$\left\langle \hat{\pi}_{j}(1)\right\rangle $ & $0.8\pm1$ & $1.5\pm1$\tabularnewline
\midrule 
$\mathcal{G}_{M}^{(M)}(m)$ & $1.2\pm1$ & $1.6\pm1$\tabularnewline
\midrule 
$\mathcal{G}_{\{M/2,M/2\}}^{(M)}(m_{1},m_{2})$ & $1\pm1$ & $0.7\pm1$\tabularnewline
\bottomrule
\end{tabular}
\par\end{centering}
\caption{Comparisons of $Z$-statistic outputs for phase-space simulations
of classical GBS and classical counts. The binned $N_{F}=4\times10^{7}$
``fake'' patterns used in these comparisons are the same as those
used for $Z_{CI}$ and $Z_{CT}$ comparisons in the main text using
squeezing parameters $\boldsymbol{r}$ and $\boldsymbol{T}$-matrix
from data sets $65\mu m,0.15W$ and $65\mu m,1.65W$. Phase-space
simulations are performed using $E_{S}=1.2\times10^{6}$ ensembles
with squashed state inputs. The superscripts denote the laser power
used for the parameters of the test. This shows excellent agreement
between the discrete and phase-space simulations.}
\end{table}

\end{document}